\documentclass[a4paper,11pt,oneside]{article}
\pdfoutput=1
\usepackage{jcappub}

\usepackage[margin=1in]{geometry}
\usepackage{physics}
\usepackage{enumerate}
\usepackage{empheq}
\usepackage{booktabs}
\usepackage{amsmath,amssymb,amsbsy,amstext, amsthm, simplewick}
\usepackage{hyperref}
\usepackage{graphicx}
\usepackage{amsfonts}
\usepackage{color}
\usepackage{wasysym}

\usepackage{tikz}

\usepackage{colortbl}
\definecolor{summersky}{cmyk}{0.71,0.33,0,0.5}
\definecolor{flamingo}{cmyk}{0,0.51,0.71,0.5}
\definecolor{rp}{cmyk}{0.2, 1, 0.6, 0}
\definecolor{pacificblue}{cmyk}{0.95,0.3,0, 0.5}
\definecolor{gray60}{cmyk}{0.4,0.4,0,0.8}

\hypersetup{
    pdftoolbar=true,        
    pdfmenubar=true,        
    pdffitwindow=true,     
    pdfstartview={FitH},    
    pdfnewwindow=true,      
    colorlinks=true,       
    linkcolor=pacificblue,          
    citecolor=flamingo,        
    filecolor=magenta,      
    urlcolor=pacificblue           
}

\newcommand{\ex}[1]{\langle #1 \rangle}
\newcommand{\be}{\begin{eqnarray} }
\newcommand{\ee}{\end{eqnarray} }
\newcommand{\bs}{\begin{split} }
\newcommand{\es}{\end{split} }

\newcommand{\vk}{\mathbf{k} }

\newcommand{\vx}{\mathbf{x} }
\renewcommand{\v}[1]{\mathbf{#1} }

\renewcommand{\L}{\mathcal{L}}

\newcommand{\B}{\mathcal{B}}

\newcommand{\Mpl}{M_{\mathrm Pl}}
\newcommand{\e}{\epsilon}

\newcommand{\then}{\quad \Rightarrow\quad}
\renewcommand{\O}{\mathcal{O}}
\newcommand{\floor}[1]{\left \lfloor{#1}\right \rfloor }

\title{\centering Building a Boostless Bootstrap\\ for the Bispectrum}

\author{Enrico Pajer}

\affiliation{Department of Applied Mathematics and Theoretical Physics, Centre for Mathematical Sciences, University of Cambridge, Wilberforce Road, Cambridge CB3 0WA, UK}
\emailAdd{ep551@cam.ac.uk}

\abstract{\noindent  The observation of primordial correlators by cosmological surveys is a very promising avenue to probe high energies and the perturbative regime of quantum gravity. Hence, it is imperative that we understand how these observables are shaped by the pillars of fundamental physics, namely unitarity, locality and symmetries. To this end, we study the three-point correlators of gravitons and scalar curvature perturbations around a quasi de Sitter spacetime. We identify a set of Bootstrap Rules that fully fix the form of these correlators in the asymptotic future, i.e. at the ``boundary'', and make no reference to ``bulk'' time evolution. Importantly, our Boostless Bootstrap accounts for the ubiquitous (spontaneous) breaking of de Sitter boosts caused by any inflationary background. We show how all bispectra involving gravitons in single-clock, canonical inflation can be easily derived in this approach. We also derive for the first time the scalar bispectrum in the Effective Field Theory of inflation to \textit{any} order in derivatives. In many cases, our derivation is computationally simpler than the corresponding explicit calculation, and makes particularly transparent the implications of locality, the choice of vacuum, and the underlying symmetries.}

\begin{document}
\maketitle
\flushbottom


\section{Introduction}


Because of the expansion history of our observable universe, the cosmological perturbations we observe must be of primordial origin. This provides a unique opportunity to study physics beyond the standard model, the quantum dynamics in curved spacetime and the perturbative regime of quantum gravity. In quantum theories of gravity, the only well-defined, gauge invariant observables are associated with the (conformal) boundary of spacetime. For Minkowski and Anti-de Sitter spacetimes, these are scattering amplitudes and conformal field theory (CFT) correlators, respectively. For an expanding, accelerated FLRW spacetime, as appropriate to model inflation, the relevant observables are equal-time correlators in the asymptotic future. The simplest correlator, namely the two-point function or power spectrum of curvature perturbations has been measured and strongly constrained by cosmological observations, while higher-point correlators, sensitive to primordial non-Gaussianity, as well as correlators involving tensor modes are the main targets of current and future cosmological surveys \cite{Meerburg:2019qqi}. Because of their theoretical and phenomenological relevance, a better understanding of correlators is one of the primary goals of much research in cosmology in the 21st century.\\

In much the same way that perturbative calculations in QFT obscure some of the structure and simplicity of scattering amplitudes \cite{Benincasa:2013,Elvang:2013cua,TASI}, the standard in-in formalism employed to compute correlators in cosmology fails to take advantage of some powerful general properties of the final result. The goal of this paper is to develop an alternative way to arrive at correlators. In particular, we shall identify a series of properties that correlators must obey and show that these are so constraining that they completely fix the result. Our focus here will be on tree-level three-point correlators, or \textit{bispectra}, because these are particularly relevant phenomenologically and particularly simple to study. We will be able to reproduce many results already known in the literature as well as derive new ones. \\

We will adopt the same general philosophy as that of on-shell methods for amplitudes, an intellectual descendant of the S-matrix program of the 60's \cite{Eden:1966dnq}. Namely, we will focus exclusively on the well-defined observables, i.e. the correlators at the future ``boundary'', and avoid discussing the un-observable time-dependent ``bulk'' dynamics that leads to these correlators. What in this approach substitutes the direct calculations of the in-in formalism are symmetries as well as general principles such as locality and unitarity. In fact, it has been known for quite some time that in de Sitter spacetime many correlators are completely fixed by symmetries. For example, in \cite{Maldacena:2011nz} it was shown that the isometries of de Sitter fix the non-perturbative graviton bispectrum. A similar result was found in \cite{Creminelli:2011mw} for the bispectrum of a spectator scalar (see also \cite{Kehagias:2012pd}). Approximate versions of de Sitter isometries were also shown to fix mixed correlators \cite{Mata:2012bx,Ghosh:2014kba,Kundu:2014gxa,Kundu:2015xta}, as well as the leading curvature bispectrum in single-clock slow-roll inflation in the limit of negligible tensor modes \cite{Pajer:2016ieg}.
More recently, a systematic formalism to compute four-particle correlators or trispectra in the presence of de Sitter isometries has been developed in a nice series of papers \cite{Arkani-Hamed:2018kmz,Baumann:2019oyu,Sleight:2019mgd,Sleight:2019hfp,Hillman:2019wgh,Baumann:2020dch} and dubbed the ``cosmological bootstrap''. In particular, building on \cite{Arkani-Hamed:2015bza}, this work has emphasized the role of unitarity, locality and the choice of vacuum in constraining the resulting correlators. \\

Given these results, it is clear that de Sitter invariant correlators are very constrained by symmetries, in an analogous way to how CFT correlators and amplitudes are restricted by the conformal and Poincar\'e symmetries respectively. However, it is much less clear how one can develop a \textit{``Boostless'' Bootstrap}, which does not rely on the full set of de Sitter isometries. This is in fact a crucial point. On the one hand, we have no observational evidence that de Sitter boosts are a good symmetry of primordial correlators because they act trivially on the only observable we have measured so far, the power spectrum. On the other hand, \textit{all} models of inflation (and in fact all cosmological backgrounds) break de Sitter boosts! This is because the background foliates spacetime with approximately homogeneous and isotropic spacelike hypersurfaces. This breaking of boosts can be small, as in canonical slow-roll models or very large, as in all single-clock models that generate a phenomenologically interesting level of primordial non-Gaussianity. To see this, recall that, assuming scale invariance, large non-Gaussianity requires a sub-luminal speed of sound, $  c_{s}\ll 1 $, which separates the sound cone from the light cone, hence breaking boosts. This observation is a particular case of a more general theorem, recently proven in \cite{Green:2020ebl}, which states that the only single-clock theory of curvature perturbations that displays full de Sitter invariance in the slow-roll, decoupling limit is a free theory. In this work we will therefore see how far we can go without ever invoking de Sitter boosts. We will see that the results for the bispectrum are very encouraging.\\

In the rest of the paper, we will identify a set of \textit{Bootstrap Rules} for the bispectrum of massless scalar and spin-two fields that become constant in the asymptotic future. Some of these rules are already implicit in the previous literature, while others are new. Some Bootstrap Rules will be derived directly from first principles, others will be deduced or inferred from explicit calculations in the in-in formalism. This is somewhat similar to the historical development of on-shell methods for amplitudes, where the simplicity of certain results was first observed at the end of a lengthy explicit calculation and only later understood in more fundamental terms (the poster child being the Parke-Taylor amplitudes \cite{ParkeTaylor}). Because of this, many of the results presented here draw from the large set of theoretical data available in the literature and the progress recently achieved in direct calculations \cite{Arkani-Hamed:2015bza,Arkani-Hamed:2017fdk,Arkani-Hamed:2018bjr,Benincasa:2018ssx,Benincasa:2019vqr,Hillman:2019wgh}. \\

All of our examples will involve only spin-0 and spin-2 fields. We expect our approach to generalize to other spins but we have not checked that in detail. Our approach can also be extended to higher-point functions and many of the principles invoked and discussed in the following have a broader validity than just the bispectrum. However, for higher-point functions additional input is certainly needed. For example, there is clearly more structure to be explored in the trispectrum because of the existence of exchange diagrams, which do not arise in the (tree-level) bispectrum. As we will discuss elsewhere, the recently derived Cosmological Optical Theorem \cite{COT} provides the necessary information to extend the current Boostless Bootstrap beyond the bispectrum.  \\

The rest of the paper is organized as follows. In the next section we summarize our main results. In Section \ref{sec:rules} we identify and discuss the Bootstrap Rules, namely covariance under rotations, translations and scale invariance (Rule 1 in Section \ref{rule1}), tree-level bispectrum in de Sitter (Rule 2 in Section \ref{rule2}), the amplitude limit (Rule 3 in Section \ref{rule3}), Bose symmetry (Rule 4 in Section \ref{ssec:rule4}), locality and the Bunch-Davies vacuum (Rule 5 in Section \ref{rule5}) and soft limits (Rule 6 in Section \ref{rule6}). In Section \ref{bootstrap} we will use the Bootstrap Rules to derive all correlators involving curvature perturbations $  \zeta $ and tensor modes $  \gamma_{ij} $ in single-field canonical inflation, namely $  \ex{\gamma\gamma\gamma} $ in Section \ref{ggg}, $  \ex{\gamma\zeta\zeta} $ in Section \ref{ssec:gzz}, $  \ex{\zeta\zeta\zeta} $ in Section \ref{ssec:scalar} and finally $  \ex{\zeta\gamma\gamma} $ in Section \ref{ssec:ggz}. We conclude with a discussion and an outlook in Section \ref{sec:conc}.

 
\subsection{The Bootstrap Rules}\label{ssec:}

%

For the convenience of the reader, we summarize in the following the Bootstrap Rules. All the rules are formulated on the boundary in terms of the actual observables, namely the future asymptotic of equal-time correlators. Notice that invariance of the correlators under de Sitter boosts is not assumed.
\begin{itemize}
\item \textit{Rule 1:} Because of homogeneity, isotropy and scale invariance, the bispectrum for fields of any spin can be decomposed into a polarization factor and a trimmed bispectrum $  \B $,
\begin{align}
B&=\sum_{\text{contractions}} \left[ \e^{h_{1}}(\vk_{1})\e^{h_{2}}(\vk_{2})\e^{h_{3}}(\vk_{3})  \vk_{1}^{\alpha_{1}}\vk_{2}^{\alpha_{2}}\vk_{3}^{\alpha_{3}}  \right]\times \B\,,
\end{align}
with index contractions left implicit. For $  \alpha_{\text{tot}}=\alpha_{1}+\alpha_{2}+\alpha_{3} $, the trimmed bispectrum must satisfy
\begin{align}
\B&=\B(k_{1},k_{2},k_{3})\,, & \B(\lambda k_{1},\lambda k_{2},\lambda k_{3})&=\lambda^{-\left( 6+\alpha_{\text{tot}} \right)} \B(k_{1},k_{2},k_{3})\,.
\end{align}
\item \textit{Rule 2:} At tree level around quasi de Sitter spacetime, for fields with an arbitrary but constant speed of sound $  c_{s} $, the trimmed bispectrum $  \B $ must be a rational function of the norms $  k_{1} $, $  k_{2} $ and $  k_{3} $ of the momenta\footnote{The exception are potential logarithms, which are discussed in Section \ref{ssec:scalar}.},
\begin{align}\label{poly}
\B=\frac{\text{Poly}_{\beta}(k_{1},k_{2},k_{3})}{\text{Poly}_{6+\alpha_{\text{tot}}+\beta}(k_{1},k_{2},k_{3})}\,,
\end{align}
for some non-negative integer $  \beta $. 
\item \textit{Rule 3:} For any $ n  $-point correlator $  B_{n} $, the residue of the total energy pole $  k_{T}\equiv \sum_{a=1}^{n}k_{a} \to 0$ is related to a corresponding amplitude $  A_{n} $ by \cite{Maldacena:2011nz,Raju:2012zr,COT}
\begin{align}\label{tep}
\lim_{k_{T}\to 0} B_{n} =\frac{(-1)^{n}H^{p+n-1}(p-1)!}{2^{n-1}} \times \frac{\Re \left( i^{1+n+p} A_{n} \right)}{\left( \prod_{a=1}^{n} k_{a} \right)^{2}k_{T}^{p}}\,.
\end{align}  
Here the order of the pole $  p $ must be positive (see \eqref{special} for $  p=0 $) and is related to the mass dimensions $  D_{\alpha} $ of the interactions responsible for the correlator by the expression
\begin{align}\label{phere}
p=1+\sum_{\alpha}(D_{\alpha}-4)\,.
\end{align}
\item \textit{Rule 4:} For the bispectrum of three identical fields, the trimmed bispectrum must be symmetric under permutations if the polarization factor is. Any symmetric polynomial can be written in a unique way in terms of sums and products of Elementary Symmetric Polynomials (ESP). Hence
\begin{align}
\B_{XXX}=\frac{\text{Poly}_{\beta}(k_{T},e_{2},e_{3})}{\text{Poly}_{6+\alpha_{\text{tot}}+\beta}(k_{T},e_{2},e_{3})}\,,
\end{align}
where
\begin{align}\label{sympoly}
e_{1}&\equiv \sum_{a=1}^{3}k_{a} =k_{T} \,,& e_{2}& \equiv \sum_{a< b}^{3} k_{a}k_{b}\,,&  e_{3}&\equiv k_{1}k_{2}k_{3}\,.
\end{align}
\item \textit{Rule 5:} Locality and the choice of the Bunch-Davies vacuum restrict the denominator of the trimmed bispectrum to take the following form
\begin{align}\label{rule4}
\B_{XYZ}=\frac{\text{Poly}_{3m+p-6-\alpha_{\text{tot}}}(k_{1},k_{2},k_{3})}{k_{T}^{p}e_{3}^{m}}\,.
\end{align}
where $  p $ is determined by \eqref{phere} and in general $  m\geq 3 $. For the symmetry breaking pattern of the Effective Field Theory (EFT) of inflation \cite{Creminelli:2006xe,Cheung:2007st}, one has $  m=3 $. We argue that a necessary condition for \textit{locality} is simply
\begin{align}\label{isoloc}
\lim_{q\to 0}\frac{B_{n}(\v{q},\v{k}_{1},\dots,\v{k}_{n})}{P(q)}< \infty\,,
\end{align}
where $  B_{n} $ is an $  n $-point correlator and $  P(q) $ is the power spectrum of the soft field.
\item \textit{Rule 6:} In single-clock cosmologies, correlators of curvature perturbations and gravitons are \textit{defined} by the soft theorems they must obey. Explicit expressions for soft scalars and soft gravitons are given in \eqref{Mcr} and \eqref{softg}, respectively.
\end{itemize}

 
\subsection{Summary of the results}\label{ssec:}

Using these rules we were able to obtain the following results
\begin{itemize}
\item The $  \ex{\gamma\gamma\gamma} $ and $  \ex{\gamma\zeta\zeta} $ bispectra in canonical slow-roll inflation follow straightforwardly from the Bootstrap Rules above (quadratic order in derivatives).
\item We re-derived the $  \ex{\zeta\zeta\zeta} $ bispectrum in canonical single-field inflation, discussing how its associated amplitude breaks boosts at $  \O(\e) $. In two cases, we were able to fix all but a part of one free coefficient in the bootstrap Ansatz: (i) at finite  $  \e $ and to leading order in slow roll and (ii) in the limit $  \e\to 0 $ and to next-to-leading order (NLO) in slow roll. In the more general non-canonical model these are indeed free parameters.
\item We derived the $  \ex{\zeta\zeta\zeta} $ bispectrum to any order in derivatives in the EFT of inflation, \eqref{any}. As an example, we discuss explicitly the case of up to cubic order in derivatives, \eqref{any3}.
\item We re-derived the $  \ex{\zeta\gamma\gamma} $ bispectrum in canonical slow-roll inflation. To this end we made use of an ad-hoc model of the flat-space amplitude.
\item We emphasize that the residue of the total-energy pole in $  \ex{\zeta\gamma\gamma} $ and $  \ex{\zeta\zeta\zeta} $ is a not manifestly local amplitude, and indeed has inverse powers of momenta. We trace this back to the existence of constrained fields in the Lagrangian descriptions, as required by gauge invariance. Ultimately, the presence of these not manifestly local terms is probably required by locality when coupling to massless spinning fields on curved spacetime.
\end{itemize}


\paragraph{Notation and conventions} We use mostly-plus signature for the metric, $  (-,+,+,+) $. Our definition of slow-roll parameters are as follows
\begin{align}\label{sr}
\e &\equiv -\frac{\dot H}{H^{2}}\,, & \eta &\equiv \frac{\dot \e}{\e H}\,, & \xi &\equiv \frac{\dot \eta}{\eta H}\,, & s&\equiv \frac{\dot c_{s}}{c_{s} H}\,,
\end{align}
where a dot denotes derivative with respect to cosmological time $  t $ and a prime a derivative with respect to conformal time, $  a d\tau \equiv dt $ with $  a $ the FLRW scale factor and $  H\equiv \dot a/a $ the Hubble parameter.


\section{The Bootstrap Rules}\label{sec:rules}

In this section, we will outline and discuss a set of Bootstrap Rules that constrain the form of the bispectra of scalar and tensor fields. These rules apply to the three-point correlators of scalar and tensor fluctuations of massless fields around a background that is well approximated by de Sitter spacetime, 
\begin{align}
ds^{2}=\frac{-d\tau^{2}+d\vx^{2}}{H^{2}\tau^{2}}\,.
\end{align}
Any flat FLRW spacetime, and in particular the flat slicing of de Sitter spacetime, is invariant under rotations and translations, which form the 3-dimensional Euclidean group ISO(3). Assuming that these isometries are also symmetries of the theory, we can classify fields according to the irreducible representations (irreps) of this group (see Appendix \ref{irreps}). Spatial translations are straightforwardly diagonalized working with Fourier space fields $  X(\vk,\eta) $ (indices on $  X $ are left implicit). At finite momentum, $  k \equiv |\vk|\neq 0 $, one can choose to work with fields that are eigenvectors of rotations by $  \theta $ around $  \vk $ with eigenvalue $  e^{i h \theta} $, where for bosons $  h  $ is an integer known as the helicity. If the theory is invariant under parity (point inversion), both $  h $ and $  -h $ fields must exist. These definite-helicity fields are represented by tensors $  X_{i_{1}\dots i_{h}} $ that are totally-symmetric, transverse and traceless over all their indices. As it is customary in dealing with cosmological tensor modes, it will be convenient to write these fields in terms of some polarization tensors\footnote{It would be much nicer to use the spinor-helicity formalism, as it's well known that it makes the analytic structure of amplitudes explicit. Indeed the spinor helicity formalism was adapted to cosmology in \cite{Maldacena:2011nz} and has been recently used to study boost-breaking theories \cite{Pajer:2020wnj}, with applications to correlators in mind (see also \cite{Fazio:2019iit}). However, due to the lack of energy conservation in cosmology, the familiar spinor-helicity expressions need to be modified. A detailed discussion will appear elsewhere.}
\begin{align}\label{def22}
X_{i_{1}\dots i_{h}}(\vk,\eta)=\e_{i_{1}\dots i_{h}}^{h}(\vk) X^{h}(\vk,\eta)+\e_{i_{1}\dots i_{h}}^{-h}(\vk) X^{-h}(\vk,\eta)\,.
\end{align}
For scalar fields we have simply $ \e^{0}(\vk) =1   $. For gravitons, $  h=\pm 2 $, we have the following useful properties
\begin{align}\label{pol1}
\e_{ii}^{h}(\v{k})&=k^{i}\e^{h}_{ij}(\v{k})=0 & \text{(transverse and traceless)}\,,\\
\e_{ij}^{h}(\v{k})&=\e_{ji}^{h}(\v{k})& \text{(symmetric)} \,, \\
\e_{ij}^{h}(\v{k})\e_{jk}^{h}(\v{k})&=0 & \text{(lightlike)} \,, \\
\e^{h}_{ij}(\v{k})\e^{h'}_{ij}(\v{k})^{\ast}&=2\delta_{hh'} & \text{(normalization)}\,, \label{normeps}  \\
\e_{ij}^{h}(\v{k})^{\ast}&=\e_{ij}^{h}(-\v{k}) &\text{($  \gamma_{ij}(x) $ is real)}  \,.\label{poln}
\end{align} 
In the following, we will compute equal-time correlators of (usually three of) these fields in the asymptotic future
\begin{align}\label{bis}
\lim_{\tau \to 0^{-}}\ex{X^{h_{1}}(\vk_{1},\tau)Y^{h_{2}}(\vk_{2},\tau)Z^{h_{3}}(\vk_{3},\tau)}=(2\pi)^{3}\delta_{D}\left(  \sum_{a}\vk_{a}\right) B_{XYZ}(\vk_{1},\vk_{2},\vk_{3},h_{1},h_{2},h_{3})\,,
\end{align}
We will sometimes leave the helicity and field-type dependence of $  B $ implicit when no confusion arises. We will denote a generic spectator scalar by $  \phi $, while we will use $  \zeta $ for curvature perturbations on constant-energy hyper-surfaces. We assume that  perturbations become constant in the asymptotic future $ \tau \to 0^{-} $, as it is known to be the case for $  \zeta $ in single-clock\footnote{By single-clock we mean models that satisfy all the soft theorems \cite{Hinterbichler:2013dpa,Creminelli:2012ed,Pajer:2017hmb,Assassi:2012zq,Hinterbichler:2012nm} that generalize Maldacena's consistency relation \cite{Maldacena:2002vr}. This definition in particular \textit{excludes} non-attractor models such as ultra-slow-roll inflation \cite{Namjoo:2012aa,Chen:2013aj}, for which different soft theorems can be derived \cite{Finelli:2017fml} (see also \cite{Bravo:2017wyw} and the mututally divergent claims in \cite{Cai:2017bxr} and \cite{Bravo:2020hde}).} inflation and for tensor modes in a wide class of models.


\subsection{Rule 1: Rotations, translations and scale invariance} \label{rule1}

By rotation invariance, all bispectra must be written in terms of contractions of the three momenta and the polarization tensors with the Kronecker delta $  \delta_{ij} $. Since the polarization tensors must appear linearly, the most general form is
\begin{align}\label{trimmed}
B&=\sum_{\text{contractions}} \left[ \e^{h_{1}}(\vk_{1})\e^{h_{2}}(\vk_{2})\e^{h_{3}}(\vk_{3})  \vk_{1}^{\alpha_{1}}\vk_{2}^{\alpha_{2}}\vk_{3}^{\alpha_{3}}  \right]\times \B \\
&=\sum_{\text{contractions}} \left( \text{polarization factor} \right) \times \left( \text{trimmed bispectrum}\right)\,,
\end{align}
where the contractions of all spatial indices in the \textit{polarization factor} (the square brackets in the first line) are implicit, the sum is over all relevant contractions with different non-negative integer powers $  \alpha_{1,2,3} $ and the \textit{trimmed bispectrum} $  \B $ for each contraction is defined by this expression. For scalars $  h_{1,2,3}=\alpha_{1,2,3}=0 $ and so $  \B $ coincides with the full bispectrum $  B $. However this is not the case in the presence of spinning particles. 

For a generic $  n $-point correlation function, $  \B $ must be invariant under rotations and translations and therefore can only depend on $  3n-6 $ independent, rotation-invariant combinations of the $  n $ momenta, for $  n\geq 3 $. A useful simplification takes place for the bispectrum, $  n=3 $, for which the three variables can be chosen to be the norms of the momenta, $  \{k_{1},k_{2},k_{3}\} $. Conversely, all higher-point correlators, with $  n\geq 4 $, depend also on some of the angles between different momenta. This feature of $  n=3 $ makes it particularly easy to impose Bose symmetry, as we will see in Rule 3. Scale invariance implies that $  B $ must be a homogeneous function of $  \{\vk_{1},\vk_{2},\vk_{3}\} $ of degree $  -6 $. In models of inflation, this scaling generally receives small slow-roll corrections, typically of order $  \e $ and $  \eta $, where
\begin{align}
\e&\equiv -\frac{\dot H}{H^{2}}\,,& \eta&\equiv \frac{\dot \e}{H\e}\,.
\end{align}
We will not attempt to recover these small corrections to the scaling exponents, but we will recover all bispectra in which the slow-roll parameters appear as overall prefactors. The reason is twofold. First, these corrections are very small in most models, and might be challenging to detecte observationally (see e.g. \cite{Sefusatti:2009xu,Byrnes:2009pe,Byrnes:2010ft,Becker:2012yr,Becker:2012je,Akrami:2019izv,Dai:2019tjh}), with some exceptions such as for example resonant non-Gaussianity \cite{Chen:2006xjb,Flauger:2009ab,Flauger:2010ja,Leblond:2010yq,Behbahani:2011it,Behbahani:2012be,CyrRacine:2011rx,Cabass:2018roz}. Second, in single-clock inflation, scale invariance can only be an exact symmetry if $  n $-point correlators scale exactly as $  k^{-3(n-1)} $. As shown in \cite{Green:2020ebl}, this can be proven directly at the level of the algebra for any correlator. Alternatively for the bispectrum one can simply notice that, if we denote by $  \Delta_{\zeta} $ the scaling dimension of $  \zeta $, scale invariance combined with the soft theorems require
\begin{align}
 \frac{1}{k^{6-3\Delta_{\zeta}}}\propto B_{\zeta\zeta\zeta} \overset{!}{=} P_{\zeta}^{2} \propto \left( \frac{1}{k^{3-2\Delta_{\zeta}}} \right)^{2}\,,
\end{align}
for which the only solution is $  \Delta_{\zeta} =0 $. 

Scale invariance can also be understood from a bulk perspective, i.e. considering the evolution of the fields as dictated by some bulk action. From this point of view, scale invariance is a consequence of the existence of an unbroken diagonal combination of time translations and some internal symmetry, both of which are individually broken (for detailed discussion see e.g. \cite{Finelli:2017fml,Nicolis:2011pv,Creminelli:2012xb,Finelli:2018upr,Baumann:2019ghk}). When combined with the assumption that the fields become constant in time in the asymptotic future, this symmetry implies the scale invariance discussed above. \\

Before concluding it is worth mentioning discrete symmetries. Since we will only be discussing neutral bosons and time is out of the picture, the only relevant discrete symmetry is parity (point inversion), $  \v{k}\to -\v{k} $. For the bispectrum of scalars, invariance under parity follows from invariance under rotations. In particular, since all three-vectors in a bispectrum must lie on the same plane by momentum conservation, a rotation $  U $ by $  180^{\circ} $ around the axis perpendicular to the plane inverts the direction of all momenta, just as parity would:
\begin{align}
\ex{\phi(\v{k}_{1})\phi(\v{k}_{2})\phi(\v{k}_{3})}&=\ex{UU^{-1}\phi(\v{k}_{1})UU^{-1}\phi(\v{k}_{2})UU^{-1}\phi(\v{k}_{3})UU^{-1}}\\
&=\ex{\phi(U\v{k}_{1}) \phi(U\v{k}_{2}) \phi(U\v{k}_{3})}\\
&=\ex{\phi(-\v{k}_{1}) \phi(-\v{k}_{2}) \phi(-\v{k}_{3})}\,,
\end{align}
where we used (with abuse of notation) that $ U\phi(\v{k})U^{-1}=\phi(U\v{k})  $ and that the vacuum is invariant under rotations. The last expression is precisely the parity transformation of the first one, and so scalar bispectra must always be invariant under parity. In passing, notice that since $  \phi(-\v{k})=\phi^{\ast}(\v{k}) $ the above equation implies that the scalar bispectrum must be real.

Conversely, spinning particles invariance under parity does \textit{not} follow from rotation invariance, and one can indeed have parity-violating bispectra. For example, for a graviton one finds the following transformations\footnote{These transformations follow from the definitions in \eqref{def22} and the fact that the graviton has two spatial indices and therefore under parity $ P\gamma_{ij}(\v{k})P=\gamma_{ij}(-\v{k}) $, while under rotations one has the standard transformation $  U\gamma_{ij}(\v{k})U^{-1}=R_{ii'}
\gamma_{i'j'}(R\v{k})(R^{-1})_{j'j} $.}
\begin{align}
P: \gamma^{\pm}(\v{k})\to P\gamma^{\pm}(\v{k})P^{-1}&=\gamma^{\mp}(-\v{k}) && \text{(parity)}\,,\\
U: \gamma^{\pm}(\v{k})\to U\gamma^{\pm}(\v{k})U^{-1}&=\gamma^{\pm}(-\v{k}) && \text{($  180^{\circ} $ rotation)}\,,
\end{align}
under parity or a rotation of $  180^{\circ} $ of the plane of the momenta, respectively. All the explicit examples in this work will be invariant under parity.\\

Summarizing, Rule 1 can be stated at the level of the trimmed bispectrum $  \B $ (defined in \eqref{trimmed})
\begin{align}
\B&=\B(k_{1},k_{2},k_{3})\,, \\
\B(\lambda k_{1},\lambda k_{2},\lambda k_{3})&=\lambda^{-\left( 6+\alpha_{\text{tot}} \right)} \B(k_{1},k_{2},k_{3})\,,\label{alphatot}
\end{align}
where $  \alpha_{\text{tot}}\equiv \alpha_{1}+\alpha_{2}+\alpha_{3} \geq 0 $. For example, for three scalars $  \alpha_{\text{tot}}=0 $.


\subsection{Rule 2: Tree-level bispectrum in quasi de Sitter spacetime} \label{rule2}

The bispectrum must be a \textit{rational function} of rotationally invariant contractions of the momenta $  \vk_{a} $ and of the possible polarization tensors $  \e^{s_{b}} $ with helicity $  s_{b}=\pm |s_{b}| $ for $  a,b=1,2,3 $.
Given the definition in \eqref{trimmed}, this means that the trimmed bispectrum $  \B $ is simply a rational function of the three norms of the momenta,
\begin{align}\label{poly}
\B=\frac{\text{Poly}_{\beta}(k_{1},k_{2},k_{3})}{\text{Poly}_{6+\alpha_{\text{tot}}+\beta}(k_{1},k_{2},k_{3})}\,,
\end{align}
where numerator and denominator are two polynomials in the norm of the momenta of degree $  \beta $ and $ 6+\alpha_{\text{tot}}+\beta $, respectively. Rule 2 is not a consequence of Rule 1, which allows for arbitrary functions of scale-invariant combinations such as $  k_{a}/k_{b} $. Rather, Rule 2 is tantamount to assuming that (i) the bulk evolution is described by a weakly coupled, local effective field theory (EFT), (ii) the free theory is well-approximated by the de Sitter mode functions for a massless fields with arbitrary but constant speed of sound $  c_{s} $, up to an arbitrary normalization\footnote{The reference normalization we have choosen corresponds to a Lagrangian $ \frac{1}{2} \left[ c_{s}^{-2}\dot \phi^{2}-(\partial_{i}\phi)^{2} \right] $.},
\begin{align}\label{modefct}
X^{h}(\vk,\tau)\propto \frac{H}{\sqrt{2 c_{s} k^{3}}} \left[ (1+i c_{s}k\tau)e^{-ic_{s} k\tau} a_{\vk}+ (1-i c_{s}k\tau)e^{+i c_{s} k\tau} a_{-\vk}^{\dagger}\right]\,.
\end{align}
Note that when $  c_{s}\neq 1 $, these mode functions are not invariant under de Sitter boosts. Intuitively this is evident in the flat-space limit where a Lorentz boost leaves the light cone invariant, but in general displaces the sound cone, unless they coincide as happens for $  c_{s}=1 $. When the mode functions deviate appreciably from the de Sitter mode functions, such as for example when the slow-roll approximation breaks down or the speed of sound has a strong time dependence, Rule 2 is generally violated (see e.g. the terms subleading in slow-roll in \cite{Chen:2006nt,Burrage:2011hd}).

As the saying goes, there's an exception to every rule. The exception to this rule is a possible logarithm of the sum of the norms, arising even when the mode functions are precisely those in \eqref{modefct}. This term features several peculiarities. First, it must be symmetric under the full group of de Sitter isometries and therefore results from the triple-$  K $ expression \cite{Bzowski:2013sza} for conformally invariant three-point functions, as shown in \cite{Pajer:2016ieg}. Second, this term is not exactly scale invariant. Instead, its variation produces a purely local term, in a way that is analogous to a conformal anomaly in a CFT. In the context of $  \zeta $ correlators in single-field inflation, the existence of this term can be thought of as a consequence of Maldacena's consistency relation when accounting for the small slow-roll deviation from the scaling $ k^{-6}  $ in the bispectrum \cite{Pajer:2016ieg}. Finally, it's worth mentioning that unitarity dictates that logarithmic terms are related to the imaginary part of the corresponding wavefunction coefficient \cite{COT}. We will discuss these logarithmic terms in Section \ref{ssec:scalar}.


\subsection{Rule 3: Amplitude limit}\label{rule3}

In the limit in which the sum of the norms of the momenta vanishes\footnote{When the fields in the correlators have different speeds of sound, the pole appears at $  \sum_{a} c_{s}^{(a)} k_{a} \to 0$. In this work, we will only consider cases where $  c_{s} $ is the same for all fields and so this reduces to the stated condition.}, 
\begin{align}
k_{T}\equiv  k_{1}+k_{2}+k_{3}  \to 0 \,,
\end{align} 
any $  n $-point correlator $  B_{n} $ is fixed by the UV-limit of a corresponding amplitude (the precise coefficient was derived in \cite{COT})
\begin{align}\label{tep}
\lim_{k_{T}\to 0} B_{n} =\frac{(-1)^{n}H^{p+n-1}(p-1)!}{2^{n-1}} \times \frac{\Re \left( i^{1+n+p} A_{n} \right)}{\left( \prod_{a=1}^{n} k_{a} \right)^{2}k_{T}^{p}}\,,
\end{align} 
for some interaction-dependent positive integer exponent $  p $ to be discussed in the following, and where we assumed canonically normalized fields for concreteness. For the bispectrum, $  n=3 $, this reduces to
\begin{align}\label{tep2}
\lim_{k_{T}\to 0} B_{3} =-\frac{H^{p+2}(p-1)!}{4} \times \frac{\Re \left( i^{p} A_{3} \right)}{\left( k_{1} k_{2} k_{3} \right)^{2}k_{T}^{p}}\,,
\end{align} 
which holds for $  p $ a positive integer. In the limiting case $  p=0 $, the correlator evaluated at time $  \tau_{0} $ possesses in general an IR logarithmic divergence\footnote{For correlators of $  \zeta $ this divergence is absent at tree-level and the $  \log {\tau_{0}} $ factor is substituted by some reference time around the time when modes exit the Hubble radius.} as $  \tau_{0}\to 0^{-} $. In this case, the same formula \eqref{tep} holds with the substitution\footnote{Note that it is only the coefficient of the $  \log\left(  -\tau_{0}k_{T}\right) $ term that can be fixed by the amplitude, but not the coefficient of $  k_{T}^{0} $. This is because the amplitude is invariant under field redefinitions while the term $  k_{T}^{0} $ is not.} \cite{COT}
\begin{align}
\frac{(p-1)!}{\left( -ik_{T} \right)^{p}}\to \log\left(  -\tau_{0}k_{T}\right)\,,
\end{align}
and so
\begin{align}\label{special}
\lim_{k_T\rightarrow0} B_3= -  \frac{ H^{2}}{(k_{1}k_{2}k_{3})^{2}}\log(-\tau_0 k_T) \Re A_3' \,.
\end{align}
These stated results deserves some explanation. A relation between cosmological correlators and amplitudes was first observed in \cite{Maldacena:2011nz,Raju:2012zr} and then discussed in \cite{Arkani-Hamed:2017fdk,Arkani-Hamed:2018kmz,Benincasa:2018ssx}. A careful proof in the in-in formalism has recently been presented in \cite{COT}, where several details of the exact form of the relation were derived that had not previously appeared in the literature. Following\footnote{Notice that \eqref{tep} is more naturally written in terms of the coefficients of the wavefunction of the universe, rather than in terms of correlators. In this paper, we avoid discussing the wavefunction of the universe because we are mostly interested in the bispectrum and the relation to the cubic wavefunction coefficient $  \psi_{3} $ is very simple
\begin{align}
P(k)&=\frac{1}{2\text{Re }\psi_{2}'(k,k)}\,, & B_{3}(\vk_{1},\vk_{2},\vk_{3})&=\frac{2\text{Re}\,\psi_{3}'(\vk_{1},\vk_{2},\vk_{3})}{\prod_{a=1}^{3}2\text{Re }\psi_{2}'(k_{a},k_{a})}\,.
\end{align}
} \cite{Arkani-Hamed:2017fdk}, we will refer to $  k_{T}=0 $ as the ``\textit{total-energy pole}''. This nomenclature originates from thinking of $  k=|\vk| $ as the energy component of a hypothetical light-like four-vector $  (k,\vk) $. From this perspective, $  k $ can be thought of as the energy associated to the wavenumber $  \vk $. In flat spacetime, when computing amplitudes (in time-translation invariant theories), we always find an energy-conserving delta function that forces the sum of all energies to vanish, and arises from a time integral of the form
\begin{align}
\int_{-\infty}^{+\infty}dt \prod_{a} e^{iE_{a}t}=2\pi \, \delta_{D}\left( E_{T} \right)\,,
\end{align}
where $  E_{T}=E_{1}+E_{2}+\dots $. Conversely, cosmological correlators do not conserve energy and are non-vanishing for generic positive values of $  k_{T} $. The reason for this is twofold. First, in cosmology we are interested in expanding spacetimes, for which time translations are spontaneously broken. This is evident for example in the form of the mode functions \eqref{modefct}, which are different from the familiar Minkowski counterpart, $  e^{ikt} $. Second, unlike amplitudes, correlators are obtained from the product of fields at some equal time, which also breaks time translations. In fact, even in Minkowski spacetime, equal-time correlators do not feature any energy conserving delta function. Instead the delta function is replaced by a pole in the total energy. For Minkowski correlators computed in the usual interaction-picture formalism, this arises from an integral over the time of the interaction, of the form
\begin{align}
\int_{-\infty}^{0} dt \prod_{a} e^{iE_{a}t}=\frac{1}{E_{T}}\,,
\end{align}
where we chose to evaluate the correlator at $  t=0 $ and we left implicit the rotation of the integral into the complex plane that projects onto the vacuum at early times and guarantees convergence. This pole at $  E_{T} $ can also be understood in a different way starting from the Lippmann-Schwinger equation, a.k.a. using old-fashioned perturbation theory (see e.g. \cite{Schwartz:2013pla}). It arises when inverting the time-independent Schr\"odinger equation to obtain the formal solution
\begin{equation}
	\ket{\psi}=\ket{\phi}+\frac{1}{E_0-H_0\pm i\varepsilon}H_{int}\ket{\psi}\,.
\end{equation}
Using this expression to compute correlators in Minkowski and setting the reference energy to zero, $  E_{0} =0$, we find that $  H_{0} $ acts on energy eigenstates to give a total energy pole $  1/E_{T} $. In cosmology, a total energy pole also arises from an integral over the conformal time $  \tau $ of the interaction, but now the order of the pole can be higher than one due to the additional powers of $  \tau $ coming from the time dependent interactions,
\begin{align}\label{schem1}
\int_{-\infty}^{0} d\tau \tau^{p-1} \prod_{a} e^{i k_{a}\tau }=\frac{1}{k_{T}^{p}}\,,
\end{align}
where again we evaluated the correlator at $  \tau = 0 $ and left the $  i\e $ prescription implicit. The special case $  p=0 $ gives the logarithmic terms we discussed in Rule 2. It would be nice to find an alternative derivation of this higher-order pole that, as in the Lipmann-Schwinger discussion for Minkowski, does not involve any time integral. \\

Using dimensional analysis and scale invariance, we can derive a useful expression for the order $  p $ of the pole for any $  n $-point correlator $  B_{n} $, defined analogously to the bispectrum $  B $ in \eqref{bis}. By dimensional analysis, in $  (3+1) $-spacetime dimensions, an $ n  $-particle amplitude and an $  n $-point correlator have mass dimension
\begin{align}\label{mdim}
[A_{n}]&=4-n\,, & [B_{n}]&=n-3(n-1)\,,
\end{align}
respectively, where we assumed that each of the fields in the correlator has mass dimension one, as it is the case for canonically normalized scalars and tensors. Notice that the amplitude is independent of the normalization of the fields, but the correlator is not. Now consider a set of interactions of dimension $  D_{\alpha} $
\begin{align}
\L \supset \sum_{\alpha} \frac{\O_{\alpha}}{\Lambda_{\alpha}^{D_{\alpha}-4}}\,,
\end{align}
where $  \Lambda_{\alpha} $ determines the corresponding coupling constant. 
For an amplitude generated by $  V $ vertices the scaling must be (at any loop order)
\begin{align}\label{there}
A_{n}\sim \frac{ E^{4-n+\sum_{\alpha}^{V} (D_{\alpha}-4) }} {\prod_{\alpha'}^{V} \Lambda_{\alpha'}^{D_{\alpha'}}}\,,
\end{align}
where $  E $ collectively represents the energy or spatial momenta of the external particles. On the other hand, scale invariance fixes the scaling with $  k $ of all (time-independent) correlators to be\footnote{The missing factors to make up the mass dimension in \eqref{mdim} are the coupling constants $  \Lambda_{\alpha} $ and factors of the Hubble parameter $  H $.}
\begin{align}\label{scaling}
B_{n}\sim \frac{1}{k^{3n-3}}\,.
\end{align}
By using the momentum/energy dependence in \eqref{there} and \eqref{scaling} in the relation \eqref{tep}, one finds
\begin{align}
E^{4-n+\sum_{\alpha}^{V} D_{\alpha} } = \frac{k^{3-3n+p+2n}}{k_{T}^{p}}\then 
p=1+\sum_{\alpha}(D_{\alpha}-4)\,.\label{master}
\end{align}
Specifying this general formula to tree-level contact diagrams we find
\begin{align}\label{master2}
p_{\alpha}=D_{\alpha}-3 \quad \quad \text{(contact interaction)}\,,
\end{align}
for any interaction of mass dimension $  D_{\alpha} $.  One can use some graph theory to rewrite this formula in an alternative way. First recall that for any graph with $  \alpha=1,\dots,V $ vertices of valence $  n_{\alpha} $ (i.e. the number legs coming out of each vertex), $  I $ internal and $  E $ external lines one has
\begin{align}
\sum_{\alpha} n_{\alpha}=2I + E\,.
\end{align}
 Then recall that the number of loops $  L $ is the first Betti number of the graph and for connected graphs it is given by
\begin{align}
L=I-V+1\,.
\end{align}
Using these relations in \eqref{master} we find the alternative expression
\begin{align}\label{gen}
p=4L+2E-3+\sum_{\alpha=1}^{V}(D_{\alpha}-2 n_{\alpha})\,.
\end{align}
In this paper we are interested in $  p $ for the tree level bispectrum, for which \eqref{master} simply tells us that \textit{$  p_{\alpha} $ is precisely the total number of time and space derivatives appearing in the interaction}
\begin{align}
p&=\text{number of derivatives} & \text{(bispectrum, $  n=3 $)}\,.
\end{align}

Summarizing, we have found that near the total energy pole, $  k_{T}\to 0 $, any correlator is fixed by a corresponding amplitude as in \eqref{tep}, with $  p $ given in \eqref{master}. In particular, this means that for contact interactions such as the tree-level bispectrum, \textit{a Laurent expansion in powers of $  k_{T} $ corresponds directly to the EFT expansion} in operators of higher and higher dimension! This observation will prove very powerful in bootstrapping the bispectrum. 


\subsection{Rule 4: Bose symmetry} \label{ssec:rule4}

Since we will consider only correlators of integer-spin fields, which must obey Bose statistics, the correlators must obey Bose symmetry when identical fields are considered. Let's begin with the case in which all three fields in the bispectrum are the same scalar. Then, the bispectrum must be symmetric under any permutation of the norms of the momenta, $  \{k_{1},k_{2},k_{3}\} $. While of course one can write a symmetric function just by starting with any non-symmetric term and summing over all permutations (``orbits''), it is very useful to make the symmetries of the problem manifest through our choice of variables\footnote{An analogous point for amplitudes has been recently made in \cite{Gripaios:2020ori,Gripaios:2020hya}. The author is thankful to W. Haddadin and especially to S. Melville for  introducing him to this branch of mathematics and explaining to him some nifty results.}. Fortunately, thanks to Rule 2 this is indeed possible.

Recall that by Rule 2 the trimmed bispectrum must be a rational function, so we can use a powerful result in commutative algebra to uniquely write down the polynomials in the numerator and denominator of \eqref{poly}. The fundamental theorem of symmetric polynomials says that any symmetric polynomial (for us on the field of real numbers) in a set of $  n $  variables can be written uniquely as sum and products of the first $  n $ elementary symmetric polynomials (ESPs)\footnote{In principle one could choose to use other families of symmetric polynomials, such as power-sum symmetric polynomials, but it turns out that the ESPs are the most convenient choice.} and numerical factors. Our interest is on $  n=3 $, in which case the relevant ESPs are
\begin{align}\label{sympoly}
e_{1}&\equiv k_{1}+k_{2}+k_{3} =k_{T}\,,\\
e_{2}& \equiv k_{1}k_{2}+k_{2}k_{3}+k_{1}k_{3}\,,\\
e_{3}&\equiv k_{1}k_{2}k_{3}\,.
\end{align}
Notice that the first ESP is precisely the total energy, and so in the following we will write $  k_{T} $ instead of $  e_{1} $. We are not the first to use ESPs to write scalar bispectra, and examples have already appeared in the literature, see e.g. \cite{Senatore:2009gt,Fergusson:2008ra}. However, it is worth stressing a few additional points:
\begin{itemize}
\item The fact that the decomposition in ESPs is unique is extremely useful to bootstrap the bispectrum, as we will see in the next section. If one tried to proceed by brute force writing down all possible monomials and summing over their permutations, one would end up with a very large number of parameters whose degeneracy cannot be constrained by any bootstrap rule.
\item ESPs are very convenient to study various limits of the bispectrum. As we discussed, the limit $  e_{1}=k_{T}\to 0 $, with $  e_{2} $ and $  e_{3} $ finite is fixed by a corresponding amplitude. Also, we will see later that the squeezed limits, which are constrained by the very powerful soft theorems, can be studied naturally by taking $  e_{3}\to 0 $ while keeping $  k_{T} $ and $  e_{2} $ finite\footnote{This leads to the natural question of what is the physical interpretation of $  e_{2}\to 0 $ with $  k_{T} $ and $  e_{3} $ finite. Unfortunately, we don't have an interesting answer at the moment.}.
\item The permutation group is particularly simple for the bispectrum, but is more complicated starting with the trispectrum because then some angles between different momenta must be kept as independent variables. Nevertheless, using results from the Hilbert series and the Hironaka decomposition one can find a complete and mutually independent set of basis polynomials, which will be presented elsewhere.
\end{itemize} 

The discussion above generalizes straightforwardly to the case in which one of the three scalars is different from the other two. The situation in which spinning fields are present is instead more interesting. Let's focus on three identical spinning fields, such as in the graviton bispectrum $  B_{\gamma\gamma\gamma} $. Now the bispectrum is given by a sum of terms, each of which is the product of a polarization factor times the trimmed bispectrum $  \B $ as in \eqref{trimmed}. When the polarization factor is permutation invariant by itself then so must be $  \B $, which can hence again be written uniquely in terms of ESPs. This is the case for example for the bispectrum generated in GR \cite{Maldacena:2002vr}. However, it might also be the case that some polarization factors are not fully permutation invariant, in which case $  \B $ is also not permutation invariant and cannot be written in terms of ESPs. This happens for example for some (higher-order) interaction induced by the coupling with the time-dependent inflation, as for the operator in $  S_{\Lambda_{3}}  $ in \cite{Bordin:2020eui}.


Summarizing, when the polarization factor in \eqref{trimmed} is invariant under permutations, the trimmed bispectrum of three identical fields can be written in terms of elementary symmetric polynomials as
\begin{align}
\B_{XXX}=\frac{\text{Poly}_{\beta}(k_{T},e_{2},e_{3})}{\text{Poly}_{6+\alpha_{\text{tot}}+\beta}(k_{T},e_{2},e_{3})}\,,
\end{align}
where $  \beta $ is a non-negative integer and $  \alpha_{\text{tot}} $ was defined below \eqref{alphatot}.


\subsection{Rule 5: Locality and the Bunch-Davies vacuum} \label{rule5}

When combined into a single fraction, the denominator of any bispectrum can only be the monomial $  k_{T}^{p} e_{3}^{m}$. This fixes the order of the numerator in terms of the integers $  p $ (discussed in Rule 3) and $  m $, 
\begin{align}\label{rule4}
\B_{XYZ}=\frac{\text{Poly}_{3m+p-6-\alpha_{\text{tot}}}(k_{1},k_{2},k_{3})}{k_{T}^{p}e_{3}^{m}}\,.
\end{align}
This rule follows from locality and the choice of the Bunch-Davies vacuum. Under the most conservative notion of locality one should choose $  m=3 $, which indeed is what appears in almost all known models. However, under a more open-minded and yet precise definition of locality one should also consider $  m>3 $, which indeed appears in solid inflation \cite{Endlich:2012pz}, where $  m=5 $. It would be nice to study that possibility further, especially given that the soft theorems for solid inflation are already known \cite{Endlich:2013jia,Pajer:2019jhb}. We leave this for the future. Let's start discussing the choice of vacuum and move on later to tackle locality.

\paragraph{Bunch-Davies vacuum} It is well known that if we assume a modified initial state, e.g. a Bogoliubov transformation of the Bunch-Davies state, we obtain additional poles in the bispectrum \cite{Chen:2006nt,Holman:2007na}. These poles can be reached even for physical configurations (i.e. with real momenta) and manifest themselves in divergences for flattened configurations, in which all three momenta become collinear $   \vk_{1} \propto \vk_{2} \propto \vk_{3} $. As pointed out in \cite{Arkani-Hamed:2015bza}, requiring the absence of these singularities is tantamount to imposing the Bunch-Davies vacuum. This fact can be seen clearly within the in-in formalism. Schematically, for contact interactions we have
\begin{align}
B\sim \Im \int_{-\infty}^{0} d\tau' \ex{H_{\text{int}}(\tau')\prod_{a} X(\vk_{a},0)}\,.
\end{align}
In the interaction Hamiltonian $  H_{\text{int}} $, only the annihilation operators contribute, which all come with the mode function $  f_{k_{a}}(\tau') $, rather than its complex conjugate. For the Bunch-Davies vacuum $  f_{k}\propto e^{-ik\tau'} $ and so the time integral over $  \tau' $ takes the form of \eqref{schem1} and gives only inverse powers of $  k_{T} $. Conversely, for modified initial states that can be parameterized as Bogoliubov transformations of the Bunch-Davies, both positive and negative frequencies appear in the mode functions $ f_{k} \sim e^{-ik\tau'} + e^{ik\tau'}   $. Then the time integral contains terms of the schematic form
\begin{align}\label{schem2}
\int_{-\infty}^{0} d\tau' \tau'^{p-1} e^{ i (k_{a}+k_{b}-k_{c})\tau' }=\frac{1}{(k_{a}+k_{b}-k_{c})^{p}}\,, \quad \quad (a\neq b\neq c)\,,
\end{align}
leading to divergences for flattened configurations $  k_{a}+k_{b}=k_{c} $ \cite{Holman:2007na}. In passing, note that the absence of these divergences was proposed in \cite{Green:2020whw} as a litmus-test for the quantum nature of cosmological perturbations.


\paragraph{Locality} Locality is a pillar of our understanding of physics and science more generally. Yet, the word locality is used in a variety of different contexts to mean different things. In the following, we will introduce a precise notion of locality for equal-time correlators and use it to constrain the form of the bispectrum.  

Let's start by clarifying our nomenclature. First, we are considering (spontaneously) boost-breaking theories, as opposed to Lorentz-invariant theories, and so the inverse spatial Laplacian $  (\partial_{i}\partial_{i})^{-1}=\partial^{-2} $ will play a key role. When writing Lagrangian interactions, we will use the expression ``manifest locality'' to denote the absence of inverse spatial Laplacians. So, for example, $ \left( \Box \phi \right)^{n}\dot\phi^{m} (\partial_{i}\phi)^{2 l} $ is a manifestly local interaction, while $  \dot \phi^{2} \partial^{-2}\dot\phi $ is not manifestly local. Manifest locality is by far too strong a restriction on the set of allowed theories. For example, already for a canonical scalar minimally coupled to general relativity we know that after solving the ADM constraints we find interactions that are not manifestly local, such as for example $  \dot \zeta^{2} \partial^{-2}\dot\zeta $ \cite{Maldacena:2002vr}. As another example, when written in terms of a canonical scalar, the interactions of a solid also have inverse Laplacians. These examples motivate us to find a more useful definition of locality. One could imagine some notion of ``bulk locality'', in which the time evolution of a given theory is studied and constraints are enforced on the propagation of signals. Here instead we will propose a definition of locality that is intrinsic to the boundary and can be defined and tested by exclusively inspecting the boundary correlators.

To gain some intuition, let's consider the simplest case of an $  (n+1) $-point scalar correlator and make the momentum-conserving delta function manifest,
\begin{align}
\ex{\phi(\v{q})\prod_{a}^{n}\phi(\v{k}_{a})}&= \int_{\v{x},\{\v{y}\}_{n}}e^{-i\left( \v{q}\cdot \v{x}+\sum_{a} \v{k}_{a}\cdot\v{y}_{a} \right)} \ex{\phi(\v{x})\prod_{a}^{n}\phi(\v{y}_{a})}\\
&= \delta_{D}^{3}\left(  \v{q}+\sum_{a}^{n}\v{k}_{a}\right) \int_{\tilde{\v{x}},\{ \tilde{\v{y}}\}_{n-1}} e^{-i \left[ \v{q}\cdot \tilde{\v{x}}+\sum_{a}^{n-1} (\v{k}_{a}-\v{k}_{n}) \cdot \tilde{\v{y}}_{a} \right]} \ex{\phi(\v{x})\prod_{a}^{n}\phi(\v{y}_{a})}\,,\label{soft}
\end{align}
where $  \tilde{\v{x}} \equiv \v{x}-\v{y}_{c} $ and similarly for $  \tilde{\v{y}} $, with $ \v{y}_{c}\equiv n^{-1}\sum_{a}^{n}\v{y}_{a} $ the ``center'' of the $  \v{y}_{a} $ locations. We can tentatively re-write the correlator by trading one field, say $  \phi(\v{x}) $, for a modification of the background (background-wave method) 
\begin{align}
\ex{\phi(\v{x})\prod_{a}^{n}\phi(\v{y}_{a})}&=\ex{\phi(\v{x}) \ex{\prod_{a}^{n}\phi(\v{y}_{a})}_{\phi} }\\
\ex{\prod_{a}^{n}\phi(\v{y}_{a})}_{\phi} &= \sum_{n=0}^{\infty} \partial^{n} \phi(\v{y}_{c}) \left[ \frac{\delta}{\delta \partial^{n}\phi(\v{y}_{c})}\ex{\prod_{a}^{n}\phi(\v{y}_{a})} \right]_{\phi=0}\,.\label{}
\end{align}
This is where locality enters: we are assuming that the product $  \prod_{a}^{n}\phi(\v{y}_{a}) $ is sensitive to the value of the background perturbation $  \phi  $ and its derivative at some nearby point $ \v{y}_{c}  $. In particular we are not allowing any dependence on, say, terms like $  \partial^{-2}\phi(\v{y}_{c}) $. Substituting the above expression in the soft limit $  \v{q} \to 0 $ of the correlator \eqref{soft}, we find 
\begin{align}
\ex{\phi(\v{q})\prod_{a}^{n}\phi(\v{k}_{a})}'&\to \int_{\tilde{\v{x}},\{ \tilde{\v{y}}\}_{n-1}} e^{-i \left[ \v{q}\cdot \tilde{\v{x}}+\sum_{a}^{n-1} (\v{k}_{a}-\v{k}_{n}) \cdot \tilde{\v{y}}_{a} \right]} \ex{\phi(\v{x})\prod_{a}^{n}\phi(\v{y}_{a})}\,, \\
&=\int_{\tilde{\v{x}}} \ex{\phi(\v{x})\phi(\v{y}_{c})} e^{-i \v{q}\tilde{\v{x}}} \times F(\v{k}_{1},\dots,\v{k}_{n})+\O(q)\\
&=P(q)\times F(\v{k}_{1},\dots,\v{k}_{n})+\O(q)\,,
\end{align}
where $  F $ is some function of the momenta $  \v{k}_{a} $ that does not depend on $  \v{q} $. This derivation tells us that \textit{locality demands that the soft limit of a correlator is controlled by the power spectrum}, namely
\begin{align}\label{isoloc}
\text{Local scalar theory} \then \lim_{q\to 0}\frac{B_{n}(\v{q},\v{k}_{1},\dots,\v{k}_{n})}{P(q)}= F(\v{k}_{1},\dots,\v{k}_{n}) < \infty\,.
\end{align}
The idea is now to use this equation, which depends exclusively on the behavior of correlators, to \textit{define} what we mean by a local theory. To do that we have to first account for spinning fields. To this end, let's switch to the (superficially) more formal language of the operator product expansion (OPE). Concentrating on the bispectrum and working in Fourier space we write 
\begin{align}\label{OPE}
\lim_{q\to 0} Y^{h_{y}}\left( \v{k}-\frac{\v{q}}{2} \right)  Z^{h_{z}} \left( -\v{k}-\frac{\v{q}}{2} \right) &\to C+\sum_{h_{x}} \left[  \e^{h_{y}}(\v{k})\e^{h_{z}}(-\v{k})\e^{h_{x}}(-\v{q})\,f(\v{k},\v{q})  \right] \, X^{h_{x}}(-\v{q})+\dots\,,
\end{align}
where $  C $ is a constant (the power spectrum of $  YZ $), and the square brackets indicate some contraction of the indices in the polarization tensors $  \e $ and in the model-dependent OPE coefficient $  f $. The dots indicate terms that are subleading when using this OPE to compute the soft-$  X $ limit of the bispectrum $  B_{XYZ} $. These are terms with higher powers of the fields, which are subleading in perturbation theory, or with more powers of $  \v{q} $, which vanish in the soft limit. The OPE limit \eqref{OPE} leads to the soft bispectrum
\begin{align}
\lim_{q\to 0} \ex{ X^{h_{x}}(\v{q})Y^{h_{y}}\left( \v{k}-\frac{\v{q}}{2} \right)  Z^{h_{z}} \left( -\v{k}-\frac{\v{q}}{2} \right)}'=P_{X^{h_{x}}}(q) \times \left[  \e^{h_{y}}(\v{k})\e^{h_{z}}(-\v{k})\e^{h_{x}}(\v{q})\,f(\v{k},\v{q})  \right]\,.
\end{align}
We now impose the condition that \textit{for a theory to be local $  f(\v{k},\v{q}) $ must be regular as $  q\to 0 $}:
\begin{align}\label{local}
\text{Local theory} \then \lim_{q\to 0}f(\v{k},\v{q})<\infty \,.
\end{align}
This is necessary condition, but it might not be sufficient. When applied to the context of this paper, this requirement of locality says that in general the denominator of the bispectrum may contain a factor of $ e_{3}^{m}  $ with $  m\geq3 $. We will mostly consider $  m=3 $ here, which is sufficient to capture non-Gaussianity in EFT of single-clock inflation, but it would be interesting to show that by allowing for $  m=5 $ one can bootstrap the bispectrum of solid inflation \cite{Endlich:2012pz}.

For a general local theory we can expand $  f $ in the soft limit as
\begin{align}
\lim_{q\to 0}f(\v{k},\v{q})=\sum_{s=0}^{\infty}\sum_{\Delta_{A}} C_{s,\Delta_{A}}(k) \left(  \frac{q}{k}\right)^{\Delta_{A}}P_{s}\left( \hat{\v{k}}\cdot \hat{\v{q}} \right)\,,
\end{align}
where $  \Delta_{A} \geq 0 $ are some exponents that must be positive\footnote{In principle $  \Delta_{A} $ could have an imaginary part, which does not change the following argument.} by locality and $  C_{s,\Delta_{A}} $ are arbitrary functions, with implicit spatial indices, which reduce to arbitrary constants over $  k^{3} $ for scale-invariant theories. Notice also that when $  Y=Z $ one must have $  C_{2s+1} =0$ by symmetry in $  \v{k}\to -\v{k} $. 

To make contact with explicit Lagrangian calculations at tree-level, notice that in the EFT of single-clock inflation, the dependence of $  f $ on $  q $ and $  \hat q $ is analytic and so the $  \Delta_{A} $'s are (non-negative) integers. Then, for interactions that are manifestly local one finds $  \Delta_{A}\geq s $ because the angular dependence only arises from $  \v{k}\cdot\v{q}=kq\,(\hat{k}\cdot\hat{q}) $. However, it is clear from our discussion that this definition of locality is too restrictive. For example, the soft limit of the scalar bispectrum of solid inflation \cite{Endlich:2012pz} has $  C_{2,0}\neq 0 $. In the OPE, such a term can only come from allowing inverse Laplacians, as in $ \left( \v{k}\cdot\v{q} \right)^{2}/(kq)^{2}  $. Despite the appearance, these theories are perfectly local according to \eqref{local}, in the sense that it is only the value of the fields and their derivatives at some point that are locally measurable. 

Furthermore, when integrating out degrees of freedom of mass $  M $, there can be additional effects on observables that are non included in the EFT of inflation because they are non-perturbative in $  M $. This results in a non-analytic dependence of $  f $ (equivalently the bispectrum) on $ k  $ and $  q $ \cite{Chen:2009zp,Baumann:2011nk,Arkani-Hamed:2015bza}. When the field that has been integrated out has non-trivial spin, this leaves a characteristic angular dependence that is referred to as cosmological collider physics in the literature \cite{Arkani-Hamed:2015bza}. This class of models obey the definition of locality introduced here as long as $  M \geq 0 $. One interesting observation is that there is connection between ``saturating'' the locality bound \eqref{local}, namely having a \textit{finite and non-vanishing} soft limit of $  f(\v{k},\v{q}) $, and the existence of interacting massless particles. We hope to come back to this issue in the future.

Finally, it's worth mentioning that higher $  n $-point correlators also display factors of $  k_{T}^{-p} $ and $  e_{n}^{m} $, but they have also other poles generated from exchange diagrams (Bulk-to-Bulk propagators in the holographic language). These singularities have been the focus of much recent interest in the literature \cite{Arkani-Hamed:2017fdk,Benincasa:2018ssx,Benincasa:2019vqr,Arkani-Hamed:2018kmz,Baumann:2019oyu,Baumann:2020dch,COT}.

Summarizing, the trimmed bispectrum must be a rational function whose denominator is fixed by locality and the choice of the Bunch-Davies vacuum to take the form of \eqref{rule4} with $  p $ given in \eqref{master} and $  m\geq 3 $ (with $  m=3 $ in the EFT of inflation).


\subsection{Rule 6: Soft limits}\label{rule6}

The rules outlined so far apply to any massless scalar or tensor fields on a (quasi) de Sitter background. However, one specific scalar and one specific tensor field are of particular interest: curvature perturbations on constant-energy hyper-surfaces $  \zeta $ and graviton perturbations $  \gamma_{ij} $. The former are the seed of the anisotropies in the cosmic microwave background and of the large scale structures in our universe and can therefore be measured from cosmological observations. Graviton perturbations have so far eluded observation but are the target of a major observational effort that will improve the current sensitivity by two orders of magnitude in the coming decades. 

When working in the bulk within a certain class of models, we carefully define $  \zeta $ and $  \gamma_{ij} $ from the fields in the action and then compute their correlators. But from an exclusively boundary perspective, where we don't assume an action to begin with, how can we know if a given correlator is the correlator of $  \zeta $ or of $  \gamma_{ij} $ as opposed to the correlator of some other scalar or tensor field? The answer to this question has been recently given in \cite{Green:2020ebl}. Within single-clock inflation and without any further assumptions about the particle content, a given correlator is a correlator of $  \zeta $ (in some cosmological model) if and only if it obeys all the soft theorems \cite{Hinterbichler:2012nm,Creminelli:2012ed,Hinterbichler:2013dpa,Pajer:2017hmb} that generalize Maldacena's consistency relation \cite{Maldacena:2002vr}. This statement is also true for correlators of $ \gamma_{ij} $ mutatis mutandis. Focusing on the bispectrum, when the momentum $  \vk_{l} $ of $  \zeta $ is much smaller than the two other short momenta, the bispectrum must obey the soft theorem
\begin{align}\label{Mcr}
\lim_{k_{l}\to 0}\ex{\zeta(\vk_{l}) X(\vk_{s}-\vk_{l}/2)Y(-\vk_{s}-\vk_{l}/2) }' = P_{\zeta}(k_{l}) \frac{\partial}{\partial \log k_{s}} \left( k^{3}\ex{X(\vk_{s} )Y(-\vk_{s})}' \right)+\dots\,,
\end{align}
where the dots stand for terms that are suppressed by at least a factor of $  \left( k_{l}/k_{s} \right)^{2} $ \cite{Maldacena:2002vr,Creminelli:2011rh,Hinterbichler:2012nm,Creminelli:2012ed} and the correct normalization for the $  \zeta $ power spectrum is the standard expression
\begin{align}
\ex{\zeta(\vk)\zeta(-\vk)}'=P_{\zeta}(k)=\frac{H^{2}}{4\e c_{s}\Mpl^{2}}\frac{1}{k^{3}}\,.
\end{align} 
There is also a soft theorem for soft gravitons, $  \gamma_{ij}(\vk_{l}) $. The leading order (LO) term in the limit $  k_{l}\to 0 $ is fixed by diff invariance to be \cite{Maldacena:2002vr,Creminelli:2012ed}
\begin{align}\label{softg}
 \ex{\gamma^{h}(\vk_{l})X(\vk_{s}-\vk_{l}/2)Y(-\vk_{s}-\vk_{l}/2)}' &\to-\frac{1}{2}P_{\gamma}(k_{l})  \e_{ij}^{h}(\vk_{l}) k_{s}^{i} \partial_{k_{s}^{j}} P_{XY}(k_{s}) \left[ 1+\O\left( \frac{k_{l}}{k_{s}} \right) \right] \,.
\end{align}
In general, diffeomorphisms cannot be used to fix the whole next-to-leading order (NLO) contribution. However, here we notice that the NLO must vanish upon averaging over the angle $  \theta $ between $  \v{k}_{l} $ and $  \v{k}_{s} $, as a consequence of parity in $  \v{k}_{l}\to -\v{k}_{l} $.

In the specific case of the bispectrum of a soft graviton and two identical fields, the full NLO soft limit is fixed to vanish\footnote{Our expression is identical to that given in \cite{Maldacena:2002vr}, but we claim it is valid also to next-to-leading order in $  k_{l} $ because we are taking the symmetric squeezed limit, which makes the next-to-leading order term vanish by symmetry. Our expression indeed agrees with asymmetric squeezed limit \cite{Creminelli:2012ed}
\begin{align}\nonumber
\ex{\gamma^{h}(\vk_{l})\zeta(\vk_{2})\zeta(\vk_{3})} &\to\frac{3}{2}\frac{H^{2}}{\Mpl^{2}k_{1}^{3}}\frac{H^{2}}{4\e\Mpl^{2} k_{2}^{3}} \frac{\e_{ij}^{h}(\vk_{l}) k_{2}^{i} k_{2}^{j}}{k_{2}^{2}}\left(  1-\frac{5}{2}\frac{\v{k}_{l}\cdot \v{k}_{2}}{k_{2}^{2}}+\dots\right)\,,
\end{align}
upon the redefinition $  \vk_{2} =\vk_{s}-\vk_{l}/2$ so that $  k_{2}\simeq k_{s} - \hat{\v{k}}_{s}\cdot \vk_{l}/2 +\dots$.}
\begin{align}\label{softg}
 \ex{\gamma^{h}(\vk_{l})X(\vk_{s}-\vk_{l}/2)X(-\vk_{s}-\vk_{l}/2)}' &\to-\frac{1}{2}P_{\gamma}(k_{l})  \e_{ij}^{h}(\vk_{l}) k_{s}^{i} \partial_{k_{s}^{j}} P_{X}(k_{s}) \left[ 1+\O\left( \frac{k_{l}^{2}}{k_{s}^{2}} \right) \right]\\
 &=\frac{3}{2} P_{\gamma}(k_{l}) \frac{\e_{ij}^{h}(\vk_{l}) k_{s}^{i} k_{s}^{j}}{k_{s}^{2}}P_{X}(k_{s})\left[ 1+\O\left( \frac{k_{l}^{2}}{k_{s}^{2}} \right) \right] \,,
\end{align}
where we assumed scale invariance, $  P_{X}\propto k_{s}^{-3} $, and the properly normalized graviton power spectrum is\footnote{One can also define the quantity $  P_{T} $ by $ P_{T}\equiv \ex{\gamma_{ij}(\vk)\gamma_{ij}(\vk')}=4P_{\gamma}  $ so that the tensor-to-scalar ratio  is $  r=P_{T}/P_{\zeta} =16\e$, at some pivot scale.}
\begin{align}
\ex{\gamma^{h}(\vk)\gamma^{h'}(\vk')}'\equiv \delta_{hh'} P_{\gamma} = \frac{H^{2}}{\Mpl^{2}}\frac{\delta_{hh'}}{k^{3}}\,.
\end{align} 
The fact that there are no NLO corrections, namely corrections at order $  k_{l}/k_{s} $, when $  X=Y $ simply follows from our symmetric way of taking the squeezed limit (see e.g. \cite{Lewis:2011au} or Appendix A of \cite{Assassi:2015jqa}) and is true both when the soft field is a scalar and a graviton. In the following, it will be useful to take the squeezed limit in expressions written in terms of the elementary symmetric polynomials. In the symmetric squeezed limit defined by 
\begin{align}\label{symsqueezed}
|\vk_{1}|&=k_{l} \,, & \vk_{2}&=\vk_{s}-\vk_{l}/2\,, & \vk_{3}&=-\vk_{s}-\vk_{l}/2\,, & k_{l}\ll k_{s}\,,
\end{align}
the leading and next-to-leading order expressions are found to be
\begin{align}
k_{T}=e_{1}&=2k_{s}+k_{l}+\O(k_{l}^{2})\,,\\
e_{2}&=k_{s}^{2}+2k_{l}k_{s}+\O(k_{l}^{2})\,,\\
e_{3}&=k_{s}^{2}k_{l}+\O(k_{l}^{3})\,.
\end{align}
Notice in particular that $ 8 e_{3}=4e_{1}e_{2}-e_{1}^{3} +  \O(k_{l}^{2}) $ and so only two variables are necessary to describe the squeezed limit at these orders\footnote{At the next order one needs a third variable, which can naturally be chosen to be $  \cos \theta_{ls}=\hat k_{l}\cdot \hat k_{s} $}. 

 
\section{Field redefinitions and boundary terms}\label{sec:}

In contrast to flat-space amplitudes, correlators are not invariant under perturbative field redefinitions and do depend on total time derivatives, both in flat and curved spacetime. In this section, we characterize these contributions to the bispectra of massless scalars, leaving a more general discussion of spinning fields and higher-point correlators for the future. The upshot will be that field redefinitions and boundary terms only contribute to the part of the correlator that is finite as $  k_{T}\to 0 $. This issue was also previously discussed in \cite{Goon:2018fyu}, with focus on manifestly local terms.


\subsection{Field redefinitions}

By field redefinitions we mean transformations of the fields in the correlator of the form
\begin{align}
X(\v{k}) \to X(\v{k})+\Delta X(\v{k})\,,
\end{align}
where $  \Delta X $ is of quadratic or higher order in the fields. Notice that these are \textit{boundary} field redefinitions, i.e. all the fields are evaluated at the boundary. From a bulk perspective, one might expect time derivatives of the fields to appear in the possible field redefinitions. However, in quasi de Sitter spacetime, these time derivatives can always be removed to leading order in (conformal) time in favor of only spatial derivatives because the equations of motion become first order near the boundary. Furthermore, not all bulk field redefinitions affect the correlators. In particular, there are field redefinitions that vanish at future infinity. For example for a scalar field, consider the following ``bulk'' field redefinitions in position and momentum space respectively,
\begin{align}
\phi(\v{x})&\to \phi(\v{x})+ c_{1} \phi(\v{x}) \bar g^{ij} \partial_{i} \partial_{j} \phi(\v{x})+ c_{2}\bar g^{ij} \partial_{i}\phi(\v{x})\partial_{j}\phi(\v{x})+\dots\\
\phi(\v{k})&\to \phi(\v{k})+H^{2}\tau^{2}\int_{\v{q}} \left[ c_{1} q^{2}+ c_{2}(\v{k}-\v{q})\cdot\v{q}\right] \phi(\v{q})\phi(\v{k}-\v{q})+\dots\,,
\end{align}  
where $  c_{1,2} $ are numerical parameters and $  \bar g^{ij} $ the background inverse de Sitter metric. Since the boundary is located at $  \tau \to 0 $, these field redefinitions are trivial at the boundary. These transformations might be useful for removing redundant operators in the Lagrangian description, as e.g. in \cite{Creminelli:2014wna,Bordin:2017hal,Bordin:2020eui}. However, they are not relevant in the present discussion, which is exclusively focused on the future boundary. Instead we will be interested in boundary field redefinitions, i.e. transformations that do not vanish as $  \tau \to 0 $. If we insist on ``\textit{manifestly local}'' field redefinitions, namely with only positive powers of derivatives, then the only possibilities are monomials in the fields. For example, for a single scalar field one has
\begin{align}
\phi(\v{x})\to\phi(\v{x})+a_{2}\phi^{2}(\v{x})+a_{3}\phi^{3}(\v{x})+\dots\,,
\end{align}
where $  a_{n} $ are numerical coefficients. Here and in the rest of this section, we focus exclusively on the momentum dependence and systematically omit the numerical normalization of the power spectrum. The only term that contributes at leading order to the scalar bispectrum is the quadratic field redefinition, which generates the much studied ``local'' non-Gaussianity \cite{Komatsu:2001rj}
\begin{align}\label{localfr}
\Delta B_{\phi^{2}}=2a_{2}\frac{\sum_{a}k_{a}^{3}}{k_{1}^{3}k_{2}^{3}k_{3}^{3}}=2a_{2}\,\frac{e_{1}^{3}-3e_{1}e_{2}+3e_{3}}{e_{3}^{3}}\,.
\end{align}
This class of field redefinitions is too restrictive though. Even for the simplest model of a canonical scalar field coupled to gravity, we should also consider ``\textit{not manifestly local}'' redefinitions, i.e. redefinitions that include inverse Laplacians (as were used in \cite{Maldacena:2002vr}). For example, the most generic redefinitions with a single inverse Laplacian to quadratic order in the scalar field is\footnote{The operator $ \partial^{2}\left(  \phi\partial^{-2}\phi \right)  $ can be re-written in terms of the others.}
\begin{align}\label{redefsnon}
\phi(\v{x}) &\to \phi(\v{x})+\partial^{-2}\left[ b_{1} \phi(\v{x}) \partial^{2} \phi(\v{x}) + b_{2}( \partial_{i}\phi(\v{x}))^{2}\right]+\\
&\quad +b_{3}\left( \partial_{i}\phi(\v{x}) \right) (\partial_{i}\partial^{-2}\phi(\v{x}))+b_{4}\left( \partial^{-2}\phi(\v{x})\right) \left(  \partial^{2}\phi(\v{x}) \right)\,,
\end{align}
where $  \partial^{2}=\partial_{i}\partial_{i} $, $  \partial^{-2} $ is the inverse spatial Laplacian and $  b_{1,2,\dots} $ are numerical coefficients. These particular combinations of derivatives have been chosen to have a net zero number of derivatives. This ensures that the induced correlators are scale invariant. From the bulk point of view, matching the number of derivatives to that of inverse Laplacians cancels the factors of $  \tau $ in the inverse metric that contracts the spatial indices, and makes these redefinitions finite at the boundary. Not all these redefinition are allowed in a local theory because of the locality condition discussed in Section \ref{rule5}. For example, let's calculate the leading contribution to the bispectrum from the redefinitions in \eqref{redefsnon}
\begin{align}\label{nonlocalfr}
\Delta B_{\partial^{-2}(\phi \partial^{2} \phi) }&= \frac{(k_{3}^{2}+k_{1}^{2})}{k_{2}^{2}k_{1}^{3}k_{3}^{3}}+\text{2 perm's}=\frac{e_{1}e_{2}-3e_{3}}{e_{3}^{3}}\,,\\
\Delta B_{\partial^{-2}(\partial_{i} \phi^{2}) }&=2 \frac{(k_{2}^{2}-k_{3}^{2}-k_{1}^{2})}{k_{2}^{2}k_{1}^{3}k_{3}^{3}}+\text{2 perm's}=2 \frac{e_{1}^{3}-4e_{1}e_{2}+6e_{3}}{e_{3}^{3}}  \,,\\
\Delta B_{ (\partial_{i}\phi)( \partial_{i}\partial^{-2}\phi)}&=\frac{1}{2 e_{3}^5} \left[ - 2 e_{1}^6 e_{3}+ e_{1}^5 e_{2}^2  + 10 e_{1}^4 e_{2} e_{3} - 5 e_{1}^3 (e_{2}^3 + 3 e_{3}^2) \right. \\
&\quad \left.- e_{1}^2 e_{2}^2 e_{3} +   4 e_{1} (e_{2}^4 + 3 e_{2} e_{3}^2) -  4 e_{2}^3 e_{3} - 6 e_{3}^3  \right] \,, \nonumber \\
\Delta B_{ (\partial^{-2}\phi) (\partial^{2}\phi)}&= \frac{e_1 e_2^4 - 4 e_1^2 e_2^2 e_3 - e_2^3 e_3 + 2 e_1^3 e_3^2 + 7 e_1 e_2 e_3^2 -  3 e_3^3}{ e_3^5} \,.
\end{align}
It is straightforward to check that only the first three of these four contributions obey the locality constraint
\begin{align}\label{localitycond}
\text{Locality:}\quad \lim_{k_{1}\to 0}\frac{B(k_{1},k_{2},k_{3})}{P(k_{1})}<\infty\,.
\end{align}
The last contribution, from the redefinition $  (\partial_{i}\phi)( \partial_{i}\partial^{-2}\phi)  $ cannot appear (by itself) in a local theory. Furthermore, the three redefinitions that contribute to order $  e_{3}^{-3} $, namely $  \phi^{2} $, $  \partial^{-2}(\partial_{i} \phi)^{2} $ and $  \partial^{-2}(\phi \partial^{2} \phi) $ are not independent from each other. They are related by
\begin{align}\label{only2}
\Delta B_{\partial^{-2}(\partial_{i} \phi)^{2} }&=\Delta B_{\phi^{2}} - 2 \Delta B_{\partial^{-2}(\phi \partial^{2} \phi)}\,.
\end{align}
Also, the redefinition $   (\partial_{i}\phi)( \partial_{i}\partial^{-2}\phi) $ contributes to the correlator at order $  e_{3}^{-5} $, rather than $  e_{3}^{-3} $. While this scaling is allowed by locality, since it obeys \eqref{localitycond}, it does not occur in models with the standard symmetry breaking pattern, namely in the EFT of inflation. We speculate that this redefinition might be useful when computing the gravity contributions to the bispectrum of solid inflation \cite{Endlich:2012pz}. In summary, there are only two independent field redefinitions that can change bispectra at order $  e_{3}^{-3} $. Remarkably, both of them were used in \cite{Maldacena:2002vr}.\\

Now let's discuss a very useful property of all field redefinitions: they only contribute to the part of the correlator that is finite in the vanishing total-energy limit, $  k_{T} \to 0$. In full generality, the reason for this is that, as we saw in Rule 3, the residue of the total-energy pole of a correlator is an amplitude and (perturbative) field redefinitions should leave amplitudes invariant. Hence field redefinitions may not have $  k_{T} $ poles. We can show this more explicitly for the bispectrum of a single scalar to leading order in the redefinition. Consider the redefinition
\begin{align}
\Delta \phi(\v{k})=\int_{\v{q}}F(\v{q},\v{k}-\v{q})\phi(\v{q})\phi(\v{k}-\v{q})\,,
\end{align}
where the kernel $  F  $ is symmetric under the exchange of its arguments, and can only depend on their scalar products. We allow $  F $ to be a rational function of momenta, but its denominator can only contain powers of $  q^{2} $, $  |\v{k}-\v{q}|^{2} $ and $  k^{2} $, to account for the presence of inverse Laplacians. The correction to the bispectrum is
\begin{align}
\Delta B=\frac{F(-\v{k}_{1},-\v{k}_{3})}{k_{1}^{3}k_{3}^{3}}+\text{2 perm's}\,.
\end{align}
This expression does not contain any pole as $  k_{T}\to 0 $, confirming our previous claim. This is of course to be expected since in the explicit bulk calculation all $  k_{T} $ poles arise from time integrals, which are absent here. 

There is an interesting corollary of this result. The most general form of the part of the scalar bispectrum that may be affected by field redefinitions is
\begin{align}
\frac{1}{e_{3}^{3}}\left[ \sum_{i,j,l \geq 0} C_{ijl} k_{T}^{i}e_{2}^{j}e_{3}^{l}\, \delta_{3, i+2j+3l} \right]=\frac{1}{e_{3}^{3}}\left[ C_{300}k_{T}^{3}+C_{110} k_{T}e_{2}+C_{001}e_{3} \right]\,,
\end{align}
for some numerical coefficients $  C_{ijl} $. Since there are three coefficients at this order but only two independent field redefinitions (see \eqref{only2}), there is one linear combination of $  C_{ijl} $ that is field-redefinition invariant, namely
\begin{align}\label{magic}
\text{Field redefinition invariant:}\quad 2 C_{300}+C_{110}+\frac{1}{3}C_{001}\,.
\end{align}
So the non-singular terms of the scalar bispectrum must contain some physical information about the interaction that can be expressed without any reference to fields. This information is in addition to the flat space amplitude, which is encoded in the residue of the $  k_{T}\to0 $ limit. An alternative perspective on this result is given by considering the wavefunction of the universe\footnote{We are thankful to Austin Joyce for discussions on this point.}. Field redefinitions correspond to contact terms in the position space wavefunction coefficients (but \textit{not} in the position space \textit{correlators}), which can be interpreted as correlators in a putative dual boundary CFT. In Fourier space, contact terms correspond to functions in which two or more momenta appear analytically. Up to permutations, there are only two possibilities, namely $  k_{1}^{3} $ and $  k_{1}k_{2}^{2} $. The function $  k_{1}k_{2}k_{3} $ may also appear but it is not analytic, and therefore cannot be changed by a field redefinition (it would correspond to correlations at separate points in the dual CFT). This indeed corresponds to the combination in \eqref{magic}. 

In summary, all field redefinitions lead to contributions to the bispectrum that are finite as $  k_{T}\to 0 $, but not viceversa: there are terms that are finite in the vanishing total-energy limit and that are invariant under field redefinitions.

 
\subsection{Boundary terms}\label{ssec:}

Interactions living on the boundary, at $  \tau =0 $, commonly arise after performing integration parts in time. These boundary terms change correlators, a fact which has been discussed several times in the literature \cite{Burrage:2011hd,Arroja:2011yj,Pajer:2016ieg}.  Here we show that contributions from boundary terms to correlators cannot diverge as $  k_{T} \to 0$. Also, we show that for every field redefinition, including those involving inverse Laplacians, we can find a boundary term that gives the same contribution to the bispectrum, a fact already noticed in \cite{Goon:2018fyu} for field redefinitions without derivatives.

Consider a boundary term in the Hamiltonian of the form
\begin{align}
H_{\partial}=\int_{\v{x}} \O(\v{x})\,,
\end{align}
for some operator $  \O(x) $, which is the product of three fields and their time or space derivatives. The contribution of this term to the correlators at leading order follows from the standard in-in formula
\begin{align}\label{forpart}
B_{3}&=i\ex{[H_{\partial},\phi(\v{k}_{1})\phi(\v{k}_{2})\phi(\v{k}_{3}) ]}'\\
&=i\int_{\v{x}} \ex{[\O(\v{x}),\phi(\v{k}_{1})\phi(\v{k}_{2})\phi(\v{k}_{3}) ]}'\,.
\end{align}
Since all fields are evaluated at the same time, namely at the boundary $  \tau=0 $, the commutator is non-zero only if at least one power of the conjugate momentum $  \Pi $ of $  \phi $ appears in $  \O $. To leading order in perturbations for a canonical scalar field this is just $  \Pi = a^{3}\dot \phi $. After using equal-time canonical commutation relations, the above expression reduces to the expectation value of the product of four fields (at equal time), which can be computed using Wick's theorem. The only non-polynomial appearances of momenta can result from inverse Laplacians in $  \O $. Since there is no time integral, poles in $  k_{T} $ cannot appear. 

In fact, for any field redefinition 
\begin{align}
\phi(\v{x}) \to \phi(\v{x})+\Delta \phi(\v{x})\,,
\end{align}
where $  \Delta \phi $ contains the product of  $  \phi $'s and their spatial derivatives, we can consider a corresponding boundary term
\begin{align}
H_{\partial}=\int_{\v{x}} \Pi(\v{x})\Delta \phi(\v{x})\,,
\end{align}
which, upon using \eqref{forpart} and canonical commutation relations gives the same contribution to correlators as the field redefinition $  \Delta \phi $.


\section{A Boostless bootstrap}\label{bootstrap}

In this section we use the Bootstrap Rules to fix the bispectra containing $  \zeta $ and $  \gamma_{ij} $ in single-clock inflation. We will be able to reproduce both the results of \cite{Maldacena:2002vr}, which assumed a canonical scalar field, as well as the bispectra that arise in more generic models captured by the Effective Field Theory of inflation \cite{Creminelli:2006xe,Cheung:2007st}, such as for example $  P(X,\phi) $ theories \cite{ArmendarizPicon:1999rj} or theories with higher derivatives.

 
\subsection{Three-graviton correlator $  \ex{\gamma\gamma\gamma} $}\label{ggg}

Let's start with the graviton bispectrum to lowest-order in derivatives, i.e. in general relativity. To determine the polarization factor in the bispectrum, we use the amplitude limit. The most general flat-space amplitude for three gravitons to lowest order in derivatives is that of GR and it is given by (see e.g. \cite{Polchinski:1998rq,Maldacena:2002vr})
\begin{align}
A_{GR}(1^{h_{1}},2^{h_{2}},3^{h_{3}})=-\frac{1}{\sqrt{2}\Mpl}\e^{h_{1}}_{ii'}(\vk_{1})\e^{h_{2}}_{jj'}(\vk_{2}) \e^{h_{3}}_{ll'} (\vk_{3}) t_{ijl}t_{i'j'l'}\,,
\end{align}
where $  t_{ijl}=k_{2}^{i}\delta_{jl}+k_{3}^{j}\delta_{il}+k_{1}^{l}\delta_{ij} $. Using the amplitude limit of Rule 3, \eqref{tep2}, this amplitude fixes the polarization factor of the correlator for this particular interaction. Now we notice that all three fields in the correlators (equivalently particles in the amplitude) are the same and that the polarization factor above is fully invariant under the permutation of any two fields, i.e. the permutation of their momenta as well as their helicities. Hence the trimmed bispectrum $  \B_{\gamma\gamma\gamma} $ can be written in terms of the three elementary symmetric polynomials discussed in Rule 4.

We now make the most generic Ansatz for the trimmed bispectrum $  \B_{\gamma\gamma\gamma} $ that is compatible with all the Bootstrap Rules. Since the GR action has only two derivatives, which is the lowest possible number for massless particles, we should take $  p=2 $. Furthermore, we should impose the correct $  k_{T} $ pole using \eqref{tep2} while adding three factors of $ \sqrt{2}/\Mpl  $ to account for the fact that the standard normalization of the graviton $  \gamma^{s} $ differs from that of a canonical scalar. The resulting Ansatz is
\begin{align}
B_{\gamma\gamma\gamma}&= \left(  \frac{\sqrt{2}}{\Mpl}\right)^{3} \frac{H^{4}}{4} \frac{\Re \left( A_{GR} \right)}{k_{T}^{2}e_{3}^{3}} \left[  e_{3}+C_{1} k_{T}e_{2} + C_{3} k_{T}^{3} \right]\\
&=-\frac{1}{2} \left(  \frac{H}{\Mpl}\right)^{4}\frac{\e^{h_{1}}_{ii'}\e^{h_{2}}_{jj'} \e^{h_{3}}_{ll'} t_{ijl}t_{i'j'l'}}{k_{T}^{2}e_{3}^{3}} \left[ e_{3}+C_{1} k_{T}e_{2} + C_{3} k_{T}^{3} \right]\,.\label{A}
\end{align}
where $  C_{1} $ and $  C_{3} $ are numerical parameters (and the label represent the corresponding power of $  k_{T} $). Notice how powerful Rules 1 through 5 are: the calculation of the graviton bispectrum has been reduced to determining $ C_{1} $ and $  C_{3} $, which are just two numbers! To fix these, we will use the (symmetric) squeezed-limit soft theorem discussed in Rule 6, namely \eqref{softg}. In the symmetric squeezed limit, we have
\begin{align}
\lim_{k_{l}\to 0}A_{\gamma\gamma\gamma}&=\e^{h_{1}}_{ii'}(\vk_{l})\e^{h_{2}}_{jj'}(\vk_{s}-\vk_{l}/2) \e^{h_{3}}_{ll'}(-\vk_{s}-\vk_{l}/2) t_{ijl}t_{i'j'l'}\\
&=2 \delta_{h_{2}h_{3}}\vk_{s}^{i}\e^{h_{1}}_{ii'}(\vk_{l}) \vk_{s}^{i'} +\dots\,.
\end{align}
Comparing the soft limit of our Ansatz \eqref{A} with \eqref{softg} to leading and next-to-leading order in $  k_{l}/k_{s} $ we find
\begin{align}
-(C_{1}+4C_{3})&=3 & 1+3C_{1}+4C_{3}=0\,,
\end{align}
which has the unique solution $  C_{1}=1 $, $  C_{3}=-1 $. So we can finally write down the graviton bispectrum as
\begin{align}
B_{\gamma\gamma\gamma}&=-\frac{1}{2} \left(  \frac{H}{\Mpl}\right)^{4}\frac{\e^{h_{1}}_{ii'}\e^{h_{2}}_{jj'} \e^{h_{3}}_{ll'} t_{ijl}t_{i'j'l'}}{k_{T}^{2}e_{3}^{3}} \left[ e_{3}+ k_{T}e_{2} - k_{T}^{3} \right] \\
&=-\frac{H^{4}}{2\Mpl^{4}} \frac{\e^{h_{1}}_{ii'}\e^{h_{2}}_{jj'} \e^{h_{3}}_{ll'} t_{ijl}t_{i'j'l'}}{k_{1}^{3}k_{2}^{3}k_{3}^{3}}\left[ \frac{k_{1}k_{2}k_{3}}{ k_{T}^{2}}+\frac{\sum_{a>b} k_{a}k_{b}}{k_{T}} -k_{T}\right]\,,
\end{align}
which indeed matches the result found in \cite{Maldacena:2002vr} via direct calculation. 

A few comments are in order. Notice that the graviton bispectrum can also be bootstrapped using exclusively de Sitter isometries, as was first done in \cite{Maldacena:2011nz}, where a second higher-derivative contribution to the graviton bispectrum was also derived. That approach is very powerful as it does not rely on perturbation theory. However, all models of inflation break de Sitter isometries via the scalar condensate. In general this breaking can be communicated to the graviton sector, although in some models such as $  P(X) $-theories this does not happen. Indeed in \cite{Creminelli:2014wna,Bordin:2017hal,Bordin:2020eui} several higher-derivative graviton bispectra were derived and studied, all of which stem from the coupling of the graviton sector to the boost-braking condensate that drives inflation and are therefore not de Sitter invariant. The bootstrap method we propose in this paper is different from (but heavily influenced by) the one in \cite{Maldacena:2011nz} and \cite{Arkani-Hamed:2018kmz}. In particular, nowhere in the above derivation did we assume invariance under de Sitter boosts. Because of this, we expect that the rules we have formulated here should be helpful to bootstrap the more general graviton bispectra that can arise from the breaking of these isometries. We leave this avenue for future work.

 
\subsection{One graviton and two scalars correlator $  \ex{\gamma\zeta\zeta} $}\label{ssec:gzz}

The next correlator that we will bootstrap is the mixed $  \gamma \zeta \zeta $ bispectrum in canonical single-field inflation, first derived in \cite{Maldacena:2002vr}. Again we start by considering the flat-space limit. In Minkowski the only amplitude corresponding to two scalars and a graviton is\footnote{This can be derived also from the spinor helicity formalism 
\begin{align}
A(1^{2},2^{0},3^{0})&\sim \frac{1}{\Mpl}\left(  \frac{[12][31]}{[23]}\right)^{2}\,, &A(1^{-2},2^{0},3^{0})&\sim \frac{1}{\Mpl}\left(  \frac{\ex{12}\ex{31}}{\ex{23}}\right)^{2}\,,
\end{align}
using the standard formulae for the polarization tensors. From this expression we immediately see that the corresponding interaction must have two derivatives, since $  [ij],\ex{ij}\sim k$. The correct normalization follows from canonically normalizing the graviton through the redefinition $  \tilde \gamma_{ij}=\Mpl \gamma_{ij} /\sqrt{2} $ applied to
\begin{align}
\L=-\frac{1}{2}\partial_{\mu}\phi \partial^{\mu}\phi-\frac{\Mpl^{2}}{8}\partial_{\mu}\gamma_{ij}\partial^{\mu}\gamma_{ij}-\frac{1}{2}\gamma_{ij}\partial_{i}\phi\partial_{j}\phi\,.
\end{align}}
\begin{align}\label{Agzz}
A_{\gamma\zeta\zeta}=A(1^{h},2^{0},3^{0})&= \frac{\sqrt{2}}{\Mpl} \e_{ij}^{h}(\vk_{1})k^{i}_{2}k^{j}_{3}\,.
\end{align}
From this we see that the corresponding cubic coupling has two derivatives, and so we should choose $  p=2 $. Naively, the boostless bootstrap would start from an Ansatz of the form
\begin{align}
B_{\gamma\zeta\zeta}& \overset{??}{\sim}  \frac{\e^{h}_{ij}k^{i}_{2}k^{j}_{3}}{k_{1}^{3}k_{2}^{3}k_{3}^{3}} \left[   \frac{e_{3}}{k_{T}^{2}} +\frac{k_{1}^{2}+k_{2}k_{3}}{k_{T}}+ k_{1}+ k_{T}\right] \,,
\end{align}
with some implicit free numerical coefficients. But this is too general. Since there are no time derivatives in the interaction, the only asymmetry between scalars and tensors appears in the polarization factor, which is fixed by the amplitude to be $   \e^{h}_{ij}k^{i}_{2}k^{j}_{3}$. Hence, we should instead start from the more restrictive ``symmetric'' Ansatz where the trimmed bispectrum is the most general polynomial of degree three written in terms of the elementary symmetric polynomials, namely
\begin{align}
B_{\gamma\zeta\zeta}&=\frac{\sqrt{2}}{\Mpl} \left(  \frac{1}{\Mpl\sqrt{2\e}}\right)^{2}\frac{H^{4}}{4}\frac{\Re A_{\gamma\zeta\zeta}}{e_{3}^{3}k_{T}^{2}}\left[   e_{3} +C_{1} e_{2} k_{T}+C_{3} k_{T}^{3}\right]\\
&= \frac{1}{4 \e } \frac{H^{4}}{\Mpl^{4}}\frac{ \e^{h}_{ij}(\v{k}_{1})k^{i}_{2}k^{j}_{3}}{e_{3}^{3}k_{T}^{2}} \left[   e_{3} +C_{1} e_{2} k_{T}+C_{3} k_{T}^{3}\right]\,,\label{symAn}
\end{align}
where the normalization was again fixed by matching the leading $  k_{T} $ pole to the flat space amplitude, as described in \eqref{tep2} while including one factor of $  \sqrt{2}/\Mpl $ to account for the non-canonical normalization of the tensor and two factors of $  1/(\Mpl\sqrt{2\e}) $ for the non-canonical normalization of $  \zeta \sim \phi/(\Mpl\sqrt{2\e}) $.

Now we can use both the scalar and the graviton soft theorems. However, both the leading and next-to-leading orders in the (symmetric) soft-scalar limit, with $  \vk_{2}=\vk_{l} $, are trivially satisfied because the polarization factor already scales as $  k_{l}^{2} $
\begin{align}
\lim_{k_{l}\to 0}B_{\gamma\zeta\zeta} &\propto \frac{ \e^{h}_{ij}(\v{k}_{s}-\vk_{l}/2) k^{i}_{l} (-k^{j}_{s} - k_{l}^{j}/2) }{e_{3}^{3}k_{T}^{2}} (2C_{1}+8C_{3})k_{s}^{3}\\
&\propto \frac{\e^{h}_{ij}(\v{k}_{s}) k^{i}_{l}  k^{j}_{l} +\O(k_{l}^{3}) }{k_{s}^{5}k_{l}^{3}}\sim \frac{1}{k_{s}^{5}k_{l}}\,,
\end{align}
where we used that the polarization tensor is perpendicular to its momentum. So all our hopes to bootstrap this correlator rest on the soft graviton theorem. This constraint indeed fixes the correlator completely. Using the definitions in \eqref{symsqueezed}, the symmetric squeezed limit must obey \eqref{softg}, namely 
\begin{align}
\lim_{k_{l}\to 0} B_{\gamma\zeta\zeta}&\overset{!}{=} - \frac{1}{2}P_{\gamma}(k_{l})  \frac{\e_{ij}^{s}(\vk_{l}) k_{s}^{i} k_{s}^{j}}{k_{s}}\frac{ \partial}{\partial k_{s}} P_{\zeta}(k_{s})\\
&=\frac{3}{2}\frac{H^{2}}{\Mpl^{2}k_{l}^{3}}\frac{H^{2}}{4\e\Mpl^{2} k_{s}^{3}} \frac{\e_{ij}^{s}(\vk_{l}) k_{s}^{i} k_{s}^{j}}{k_{s}^{2}}+\O(k_{l}^{-1})\,.
\end{align}
Taking this limit on \eqref{symAn}, we find again
\begin{align}
-(C_{1}+4C_{3})&=3\,, & 1+3C_{1}+4C_{3}=0\,,
\end{align}
which is again solved by $ C_{1}=1 $ and $ C_{3}=-1$. Replacing these values in our original Ansatz \eqref{symAn} delivers the correct result, namely \cite{Maldacena:2002vr}
\begin{align}\label{gzz}
B_{\gamma\zeta\zeta}=\frac{1}{4\e} \frac{H^{4}}{\Mpl^{4}} \frac{\e^{h}_{ij}(\vk_{1})k^{i}_{2}k^{j}_{3}}{k_{1}^{3}k_{2}^{3}k_{3}^{3}}\left[ \frac{k_{1}k_{2}k_{3}}{k_{T}^{2}}+\frac{\sum_{i>j} k_{i}k_{j}}{k_{T}} -k_{T}\right]\,.
\end{align}

A few comments are in order. This bispectrum, originally derived in \cite{Maldacena:2002vr}, was re-derived using an approximated version of de Sitter invariance in \cite{Ghosh:2014kba}. Most recently, again using de Sitter isometries as input, but adopting a more systematic formalism, this correlator was bootstrapped in \cite{Baumann:2020dch} from a related correlator involving conformally coupled fields by means of an appropriate weight-raising operator. The fact that in this section we were able to derive $  \ex{\gamma\zeta\zeta} $ without ever mentioning de Sitter boosts might be at first surprising. What's happening is that the soft-graviton consistency relation as well as the choice of a Lorentz invariant amplitude are supplying the constraining power that comes from de Sitter boosts in the other approaches. This is an important difference because, as first stressed in \cite{Hinterbichler:2012nm}, the soft theorems do not depend on the isometries of de Sitter. Rather a conformal group emerges when considering adiabatic modes \cite{Weinberg:2003sw} in any accelerating FLRW at future spacelike infinity.

 
\subsection{Three-scalar correlator $  \ex{\zeta\zeta\zeta} $}\label{ssec:scalar}

Let's consider the most phenomenologically relevant of correlators, the scalar bispectrum of curvature perturbations $  \ex{\zeta\zeta\zeta} $. A pleasant simplification occurring in this case is that the polarization factor in \eqref{trimmed} is trivial and so the full bispectrum coincides with the trimmed bispectrum. 

Before we can write down an Ansatz that respects all of the Bootstrap Rules, we need to consider what amplitudes will appear on the total-energy pole. In particular, we have to specify whether these amplitudes display linearly-realized Lorentz invariance, as is customarily assumed in the amplitude literature \cite{Benincasa:2013,Elvang:2013cua,TASI}, or if they (spontaneously) break Lorentz invariance \cite{Pajer:2020wnj} (see also \cite{Conjecture,Grall:2020tqc}). People that are used to working with particle physics in flat spacetime might think that the former is the natural choice, however this is not the case in cosmology. As is well-known, and discussed at length recently in \cite{Pajer:2020wnj}, Lorentz boosts are generically broken in cosmology. During inflation, the breaking is induced by the presence of the homogeneous and isotropic inflaton background, which picks out a preferred frame. In single-field inflation with a canonical kinetic term, the coupling of perturbations to this boost-breaking background is suppressed by the slow-roll parameter $  \epsilon $. Therefore, as proven in detail in \cite{Pajer:2016ieg}, in the limit $  \e \to 0 $, the flat spacetime limit of the theory should enjoy Lorentz invariance, which for the full correlator becomes an (approximate) conformal symmetry\footnote{By Theorem 2 of \cite{Green:2020ebl} the conformal symmetry cannot be exact or else all connected correlators would vanish. Indeed the conformal symmetry here is weakly broken by small corrections of order $  \eta $ to the scale dependence.}. Hence, to reproduce the result of Maldacena for canonical single-field inflation \cite{Maldacena:2002vr} in the limit $  \e \to 0 $, we should consider Lorentz-invariant amplitudes. 

However, there are other possibilities. It is well known that when a non-canonical kinetic term is present, such as in $  P(X,\phi) $-theories, the coupling with the boost-breaking background can be arbitrarily strong (so much that it can make the theory non-perturbative). Indeed, this is made manifest in the formalism of the Effective Field Theory of inflation \cite{Creminelli:2006xe,Cheung:2007st}, where infinitely many terms arise that are \textit{not} individually invariant under boosts. Rather, boosts are spontaneously broken and therefore non-linearly realized. To bootstrap the predictions of this more general class of theories, we should consider the recently derived boost-breaking amplitudes of \cite{Pajer:2020wnj}. The most generic (analytically continued) \textit{local} amplitude for three identical massless scalars in a perturbative boost-breaking EFT is
\begin{align}\label{A3}
A_{\phi\phi\phi}=A(1^{0},2^{0},3^{0})=F(E_{1}E_{2}+E_{1}E_{3}+E_{3}E_{2},E_{1}E_{2}E_{3})\,,
\end{align}
where $  F $ is an arbitrary polynomial of its two variables, which can be recognized as the second and third elementary symmetric polynomials (ESP's) for the three energies. When the theory is boost invariant, the function $  F $ must reduce to a constant (since there are no other non-vanishing Lorentz invariants). Let's consider these cases separately.


\subsubsection*{Boost-invariant amplitudes, a.k.a. the conformal limit of canonical inflation} 

The only local, Lorentz-invariant amplitude for three massless scalars is a constant and therefore the associated number of derivatives is zero. So we should choose $  p=0 $ in our Ansatz for the bispectrum. This limiting value for $  p $ should be interpreted as the fact that the polynomial divergence in $  k_{T}\to 0 $ becomes a logarithmic one, as in \eqref{special}. The coefficient of the logarithmic divergence is the fixed by the flat space amplitude according to \eqref{special}, where as usual we should include three factors of $  (\sqrt{2\e}\Mpl)^{-1} $ to account for the non-canonical normalization\footnote{Since we have in mind canonical inflation, we have set $  c_{s}=1 $ in this normalization.} of $  \zeta $. To make contact with the standard discussion, we can write the constant, Lorentz-invariant (LI) amplitude simply as $  V'''(\bar \phi) $ where $  V(\phi) $ is the inflaton potential. Using the background equations of motion in the limit $  \e \to 0 $, this can be conveniently written as (see e.g. \cite{Pajer:2016ieg} for a derivation)
\begin{align}
A_{\phi\phi\phi}^{LI}=\lim_{\e\to 0}V'''=- \frac{3}{2}\,\frac{H^{2}}{\sqrt{2\e}\Mpl}\, \dot \eta +\dots\,,
\end{align}
where the dots stand for terms of higher order in slow-roll, such as $  \eta^{2} $ and $  \eta \dot\eta $. Our Ansatz therefore takes the form
\begin{align}\label{zAns}
B_{\zeta\zeta\zeta}=\left(  \frac{H^{2}}{4\e \Mpl^{2}}\right)^{2}\frac{1}{e_{3}^{3}}\left[  \frac{\dot  \eta}{2H}\log\left(  -k_{T} \tau_{\ast} \right)\left(  3e_{3}+C_{1}k_{T}e_{2}+C_{3}k_{T}^{3}\right)+D_{0}e_{3}+D_{1}k_{T}e_{2}+D_{3}k_{T}^{3}\right]\,,
\end{align}
where $  C_{1,3} $ and $  D_{0,1,3} $ are five numerical parameters that we wish to determine and $  \tau_{\ast} $ is some fixed time, which without loss of generality can be chosen to be the Hubble-crossing of the shortest mode in the correlator. The soft limit \eqref{Mcr} to next-to-leading in slow-roll for $ \e \to 0 $ is
\begin{align}
\lim_{k_{l}\to 0} B_{\zeta\zeta\zeta} \left( \vk_{l},\vk_{s}-\frac{\vk_{l}}{2},-\vk_{s}-\frac{\vk_{l}}{2}\right)&=(1-n_{s})P_{\zeta}(k_{l})P_{\zeta}(k_{s})\\
&=P_{\zeta}(k_{l})P_{\zeta}(k_{s})\left[ \eta_{\ast}+\frac{\dot\eta}{H}\left(  \gamma_{E}-2+\log2\right)+\O\left( \frac{k_{l}^{2}}{k_{s}^{2}} \right) \right] \,,\nonumber
\end{align}
where $  \eta_{\ast}=  \eta(\tau_{\ast}) $ and we can safely neglect to specify the time at which $  \dot\eta $ and $  H $ are evaluated because it is higher order in the slow-roll expansion. Imposing this condition on our Ansatz, we find the two constraints
\begin{align}
D_{0} + 5 D_{1} + 12 D_{3} + 
 \frac{\dot\eta}{2H}  \left[ C_{1} + 4 C_{3} + \log(8) + C_{1} \log(32) + 2 C_{3} \log(64) \right] + \nonumber\\
+  \frac{\dot\eta}{2H} (3 + 5 C_{1} + 12 C_{3}) \log (- k_{s} \tau_{\ast}) &=0 \,,\\
 2 D_{1} + 8 D_{3} - \eta_{\ast} + \frac{\dot\eta}{H} (2 - \gamma_{E} + (-1 + C_{1} + 
       4 C_{3}) \log(2))+ \nonumber \\
       + (-1 + C_{1} + 4 C_{3}) \frac{\dot\eta}{H} \log(-k_{s} \tau_{s})&=0\,.
\end{align}
Since the terms proportional to $  \log(-k_{s} \tau_{s}) $ must vanish separately, we find the following solutions
\begin{align}\label{1}
C_{1}&=-3\,, & D_{0} &= \frac{1}{2}  \left[ -4 D_{1} + \frac{\dot\eta}{H} (5 - 3 \gamma_{E}) - 3 \eta_{\ast} \right]\,,\\
C_{3}&=1\,, & D_{3}&= \label{2}
 \frac{1}{8} \left[-2 D_{1} +\frac{\dot\eta}{H} (-2 + \gamma_{E})  + \eta_{\ast} \right]\,.
\end{align}
Notice that the Bootstrap Rules completely fix the logarithmic term. Unfortunately, in this particular setting, we were not able to fix the last parameter $  D_{1} $ without appealing to de Sitter boosts\footnote{The part of $  D_{1} $ proportional to $  \eta_{\ast} $ can actually be determined by the following argument. If we set $  \dot\eta =0 $ the correlator must come exclusively from a field redefinition and therefore must satisfy \eqref{magic}, which can be used to find $  D_{1}=-3\eta/2 $. Alternatively, one can demand that the bispectrum does not depend on the choice of time to this order in slow-roll, which gives the same result. Both of these arguments do not say anything about the part of $  D_{1} $ proportional to $  \dot\eta $.}. To account for boost invariance, the condition to be imposed is (see e.g. \cite{Bzowski:2013sza})
\begin{align}
\left( K_{a}-K_{b} \right)B_{\zeta\zeta\zeta}(k_{1},k_{2},k_{3})=0\,,
\end{align}
for $  a\neq b $, $  a,b=1,2,3 $, where
\begin{align}
K_{a}=\frac{\partial^{2}}{\partial k_{a}^{2}}+\frac{4}{k_{a}}\frac{\partial}{\partial k_{a}}\,.
\end{align}
Substituting the solutions \eqref{1} and  \eqref{2} into our Ansatz \eqref{zAns} and imposing boost invariance we find
\begin{align}
D_{1}=\frac{\dot\eta}{H} \left( 1-\frac{3}{2}\gamma_{E} \right) - \frac{3}{2}\eta_{\ast}\,.
\end{align}
This leads to the final result
\begin{align}
B_{\zeta\zeta\zeta}&=\left(  \frac{H^{2}}{4\e \Mpl^{2}}\right)^{2} \frac{1}{e_{3}^{3}} \left\{ \frac{\eta_{\ast}}{2}(k_{T}^3 - 3 k_{T} e_{2} + 3 e_{3}) + \frac{\dot\eta}{2H}(k_{T}^3 - 3 k_{T} e_{2} + 3 e_{3}) \log(-k_{T}\tau_{\ast} )\right. \\
&\hspace{3cm}+\left.\frac{\dot\eta}{2H}
  \left[ e_{3} (1+3\gamma_{E}) + e_{1} e_{2} (2 - 3 \gamma_{E}) + 
    k_{T}^3 (\gamma_{E}-1)  \right]  \right\}\\
 &= \left(  \frac{H^{2}}{4\e \Mpl^{2}}\right)^{2} \Bigg\{\frac{\eta_{\ast}}{2} \frac{k_1^3 + k_2^3+k_3^3}{k_1^3 k_2^3k_3^3}+ \frac{\dot{\eta}}{2H}\frac{1}{k_1^3k_2^3k_3^3}\times \\
& \hspace{3cm}   \Big[(-1+\gamma_E+ \log{(-k_{T} \tau_*)})\sum_{i=1}^3 k_i^3-\sum_{i\neq j}k_i^2 k_j +  k_1 k_2 k_3\Big]\Bigg\}\,, \nonumber
\end{align}
which indeed agrees with the direct calculation \cite{Zaldarriaga:2003my,Seery:2008qj,Creminelli:2011mw}\footnote{As pointed out in \cite{Seery:2008qj}, the result quoted in (25) of \cite{Zaldarriaga:2003my} missed a few terms. The problem is that the decaying part of the mode functions at the boundary cannot be dropped as it is usually done for correlators with a larger number of derivatives. See Sec 3.3 of \cite{Enriconotes} for a pedagogical derivation.} and the symmetry-based derivation in \cite{Pajer:2016ieg}.


\subsubsection*{Boost-breaking amplitudes at two derivatives: canonical inflation and locality} 

We now move on to consider boost-breaking amplitudes \cite{Pajer:2020wnj} for three scalars at quadratic order in derivatives. We begin with a discussion of the $  \zeta $ bispectrum with at most two derivatives ($  p=2 $). This will lead to canonical single-field inflation, whose bispectrum was originally derived in \cite{Maldacena:2002vr}. In the next subsection we will move on to any higher number of derivatives, $  p\geq 3 $, and re-derive and generalize the bispectrum in the EFT of inflation.

Since this case is the most complicated of all bispectra considered here, to gain some intuition we start from the solution of the explicit calculation \cite{Maldacena:2002vr}\footnote{Here we used 
\begin{align}
\frac{2\ddot \phi}{\dot \phi H}=\eta-2\e\,, \quad \e=\frac{\dot \phi^{2}}{2\Mpl^{2}H^{2}}\,.
\end{align}} 
\begin{align}
B_{\zeta\zeta\zeta}&= \frac{1}{2}\left(  \frac{H^{2}}{ 4 \e \Mpl^{2}}\right)^{2}\frac{1}{k_{1}^{3}k_{2}^{3}k_{3}^{3}} \left[    \left( \eta-2\e  \right) \sum_{a} k_{a}^{3}+\e \left( \sum_{a} k_{a}^{3}+\sum_{a\neq b}k_{a}k_{b}^{2}+ 8\frac{\sum_{a>b} k_{a}^{2}k_{b}^{2}}{ k_{T}} \right) \right]  \nonumber \\
&= \frac{1}{2}\left(  \frac{H^{2}}{ 4 \e \Mpl^{2}}\right)^{2}\left[ \frac{(\eta-\e)(k_{T}^{3}-3k_{T}e_{2}+3e_{3})}{e_{3}^{3}}+\e\frac{k_{T}e_{2}-3e_{3}}{e_{3}^{3}}+8\e\frac{e_{2}^{2}-2k_{T}e_{3}}{e_{3}^{3}k_{T}} \right]\,.\label{exact}
\end{align}
The first remarkable fact is that the leading total-energy pole is $  k_{T}^{-1} $, corresponding to $  p=1 $. We can interpret this in two distinct but related ways: the consequence of non-linearly realized boosts that forbid $  p=2 $ or as integrating out non-dynamical fields. Let's consider these two points of view in turn.

\paragraph{The absence of $  k_{T}^{-2} $ poles} The first possibility is to recall that both gravity and the canonical scalar field are theories with at most two derivatives, hence we would expect $  p=2 $. While there are certainly boost-breaking amplitudes with two derivatives (e.g. $  A_{3}\sim e_{2} $ from $  \zeta\dot\zeta^{2} $), when boosts are non-linearly realized it is not possible to have such two-derivative interactions without having also three derivative interactions, which in this subsection we are assuming to vanish. To see this, it would be nice to have an elegant on-shell derivation based on the Ward-Takahashi identities of Lorentz boosts. Instead here we limit ourselves to a more pedestrian argument. In the covariant field-theoretic formulation of the theory, the only cubic interaction with two derivatives are $  \phi (\partial_{\mu}\phi)^{2} $. But in flat space this interaction can be removed by a field redefinition and so the corresponding amplitude must vanish\footnote{The flat space limit of the EFT of inflation offers an alternative perspective. The building blocks of the EFT are $ M_{n}(t+\pi)  [2\dot\pi-(\partial_{\mu}\pi)^{2}]^{n} $ for $  n\geq 2 $. Hence, the only possibility for a two-derivative interaction comes from expanding $  M_{2} $ into $  \dot M_{2} \pi \dot \pi^{2} $. But this term comes also with interactions that have a higher number of derivatives, e.g. $  \dot M_{2} \pi \dot \pi (\partial \pi)^{2} $ or $  M_{2} \dot \pi (\partial\pi)^{2} $. There is a somewhat singular case when at some instant in time $  M_{2}=0 $ and $  \dot M_{2} \neq 0 $. At this instant one can indeed have cubic interactions that have two derivatives but not more. However interactions with three or more derivatives always appear at any other time as well as in quartic and higher interactions.}. When the amplitude vanishes, the relation \eqref{tep} tells us that the total-energy residue of the correlator is zero and so the leading total-energy pole appears at the next order (a similar phenomenon was also noticed in \cite{GJS} for DBI), which for us is $  p=1 $. In passing, this argument implies that the residue of $  k_{T}^{-2} $ in canonical single-field inflation should vanish to all orders in the slow-roll parameters. By carefully studying\footnote{There is a typo in Table 2 in the expression for $  Z_{a} $ corresponding to $  \zeta (\partial\zeta)^{2} $, the factor of 3 in front of $  c_{s}^{2} $ should be removed.} the lengthy results of \cite{Burrage:2011hd} we checked that this indeed happens at next-to-leading order in slow-roll, $  \O(\e^{2},\e \eta, \eta^{2},\eta \xi) $. 

We are now in the position to bootstrap the bispectrum starting from the following Ansatz with $  p=1 $
\begin{align}\label{zzzC}
B_{\zeta\zeta\zeta}=\frac{1}{2}\left(  \frac{H^{2}}{ 4 \e \Mpl^{2}} \right)^{2} \frac{C_{0} e_{2}^{2}+C_{1}k_{T}e_{3}+C_{2}k_{T}^{2}e_{2}+C_{4}k_{T}^{4}}{e_{3}^{3}k_{T}}\,,
\end{align}
where as usual $  C_{0,1,2,4} $ are numerical constants to be determined and without loss of generality the normalization has been chosen for future convenience. 
We can now impose Maldacena's consistency relation, \eqref{Mcr} and find
\begin{align}\label{scalarconst}
\frac{1}{4}  (C_{0} + 4 C_{2} + 16 C_{4}) &=1-n_{s}\,, & 7 C_{0} + 4 C_{1} + 20 C_{2} + 48 C_{4}&=0\,.
\end{align}
We can use these two expressions to fix two coefficients, say $  C_{2} $ and $  C_{4} $ in terms of $ 1-n_{s}$ and the other two $  C $'s. Then the result of the Boostless Bootstrap is a bispectrum with 3 free parameters, namely $  C_{0} $, $ C_{1}  $ and $  n_{s}-1 $. The value for $  C_{0} $ can be determined from the normalization of the flat space amplitude, which we will discuss in the next subsection, around \eqref{340}-\eqref{341}. From the exact result in \eqref{exact} and standard slow-roll results we see that
\begin{align}\label{expl}
C_{0}& =8\e \,, & C_{1}&=3\eta-22\e\,, &  C_{2}&=4\e -3 \eta\,, & C_{4}&= \eta - \epsilon \,,
\end{align}
which indeed satisfies \eqref{scalarconst} recalling that $   1-n_{s}=2\e+\eta $. This tells us that actually the full result has only 2 free parameters, namely $  \e $ and $  \eta $. In other words, to leading order in slow-roll there is an additional relation among the free parameters that is not captured\footnote{Actually the $  \eta $ dependence of $  C_{1} $ can be fixed. In the next subsection we will show that $  C_{0}=8\e $ by computing the normalization of the flat space amplitude. Then, we notice that as $  \e\to 0 $, $  C_{1} $ must arise exclusively from a field redefinition and therefore it must satisfy \eqref{magic}. This fixes $ \lim_{\e\to 0} C_{1}=3(1-n_{s}) $, which indeed agrees with the explicit calculation, \eqref{expl}.} by our current Bootstrap Rules. It is possible that either to higher order in the slow-roll expansion the relation found to leading order breaks down, or that we are indeed missing some important constraint, perhaps related to the non-linear realization of boosts. We do not pursue this further because all these contributions to the bispectrum end up being degenerate with others that arise when considering more general theories with non-canonical kinetic terms and higher derivatives, as captured by the EFT of inflation.

\paragraph{Exchange or contact? The strange case of constrained fields}

There is a second way to understand why the $  \zeta $ bispectrum has $  p=1 $, which highlights the peculiar behavior of gravity in curved spacetime. Indeed, we know from \cite{Ghosh:2014kba,Arkani-Hamed:2015bza} (see also the discussions in \cite{Arkani-Hamed:2018kmz,Baumann:2020dch}) that the $  \zeta $ bispectrum arises from the ``graviton-exchange'' scalar trispectrum\footnote{This is not the trispectrum from the exchange of a transverse-traceless graviton computed in the in-in formalism in \cite{Seery:2008ax}, which does not contribute to the soft trispectrum.} after ``putting a leg on the background'', i.e. after taking one of the momenta to zero and appropriately rescaling the correlator. This procedure generates the leading $  k_{T}^{-1} $ pole that appears in Maldacena's explicit computation. Following this intuition and in contrast to the alternative interpretation presented in the previous subsection, here we attempt to interpret the residue of the $  k_{T}^{-1} $ pole in \eqref{exact} as a flat space amplitude, as suggested by the Bootstrap Rule 3. 

Interpreting the $  k_{T}^{-1} $ pole in \eqref{exact} as a flat spacetime amplitude seems immediately problematic. No manifestly local perturbative three-particle amplitude exists with just one derivative! Even in the boost-breaking case, the amplitude starts quadratic in derivatives because the only symmetric linear term is $  k_{T} $, which vanishes by conservation of energy (see \eqref{A3}). So how can this interpretation be tenable? The key to the answer is that locality has different implications in flat and curved spacetime. In more detail, in the presence of a non-trivial inflaton background, gravity leads to an \textit{apparently} non-local amplitude in the flat-space limit, namely
\begin{align}\label{nonl}
A_{GR}(1^{0},2^{0},3^{0})\propto \frac{\left( E_{1}E_{2}+E_{1}E_{3}+E_{3}E_{2} \right)^{2}}{E_{1}E_{2}E_{3}}=\frac{e_{2}^{2}}{e_{3}}\,,
\end{align}
where, with an abuse of notation, we used $  e_{2} $ and $  e_{3} $ to also denote the elementary symmetric polynomials of the three energies $  E_{1,2,3} $. The expression \eqref{nonl} appears non-local because of the inverse powers of energy, $  e_{3}^{-1} $. One can check that this amplitude corresponds to the not manifestly local interaction $  \dot \zeta^{2} \partial^{-2}\dot \zeta $, which is indeed the only interaction that contributes to the bispectrum\footnote{This interaction is responsible for the last term in \eqref{exact}, while the first two terms come from field redefinitions (see \eqref{localfr} and \eqref{nonlocalfr})}.

The emergence of this apparent non-locality is clear in the explicit field theory calculation \cite{Maldacena:2002vr}, and we provide a field theory toy model in Appendix \ref{app:toy} that captures the main features of this mechanism. Equivalently, we can derive \eqref{nonl} from the four-particle amplitude in a theory with \textit{local} interactions of a massless scalar with a \textit{non-dynamical} scalar with propagator $  i/|\vk|^{2} $. The interaction should be the same as that of two scalars with a spin-2 particle, namely \eqref{Agzz}, but we should substitute the polarization tensor $  \e_{\mu\nu} $ with $  \delta_{\mu0}\delta_{\nu0} $. The four-particle amplitude is then simply dictated by factorization (see \eqref{A4app} in Appendix \ref{app:toy} for an explicit calculation)
\begin{align}\label{340}
A_{4}&= -2 \frac{E_{1}E_{2}}{\Mpl}\times \frac{i}{|\v{k}_{1}+\v{k}_{2}|^{2}}\times  \frac{E_{3}E_{4}}{\Mpl}+\text{2 perm's}\,.
\end{align}
To obtain the three-particle amplitude, we need to ``put a leg on the background''. To this end, we send one of the four-momenta to zero, say $ E_{4} , \v{k}_{4}\to 0 $, and rescale by a factor $   \dot{\bar \phi}/(iE_{4}) $ to account for the normalization of one-particle states, where the particle energy $  E_{4} $ differs from the ``background energy'' $  \dot{\bar\phi} $. We therefore obtain
\begin{align}\label{341}
A_{\varphi\varphi\varphi}&= \lim_{E_{4},\v{k}_{4}\to 0}\, \frac{\dot{\bar \phi} }{iE_{4}}A_{4}= 2i\frac{\dot{\bar \phi}}{\Mpl^{2}}E_{1}E_{2}E_{3}\left(  \frac{1}{E_{3}^{2}}+ \frac{1}{E_{2}^{2}}+ \frac{1}{E_{1}^{2}}\right)\\
&=2i \frac{\dot{\bar \phi}}{\Mpl^{2}} \frac{e_{2}^{2}}{e_{3}}=i \frac{2\sqrt{2\e}H}{\Mpl} \frac{e_{2}^{2}}{e_{3}}\,,
\end{align}
where we re-wrote $  \dot{\bar \phi} $ in terms of $  \e $. In addition to reproducing the correct scaling of the non-local amplitude, this derivation fixes the overall coefficient. In particular, the amplitude must be suppressed by the slow-roll parameter $  \e $. This argument can be used to fix the free coefficient $  C_{0} $ in \eqref{zzzC} to precisely the value found in the explicit calculation, \eqref{expl}.


\subsubsection*{Boost-breaking amplitudes to any order in derivatives: the EFT of inflation}

We can now extend our discussion to any order in derivatives. This derivation will reproduce all possible bispectra generated by all possible operators in the EFT of single-clock inflation \cite{Cheung:2007st}. 

For a theory with $  p $ derivatives, we start from the Ansatz 
\begin{align}\label{any}
B_{\zeta\zeta\zeta}&=\frac{\text{Poly}_{p+3}(k_{T},e_{2},e_{3})}{e_{3}^{3}k_{T}^{p}}\\
&=\frac{1}{e_{3}^{3}k_{T}^{p}}\sum_{i,j,l\geq 0} C_{i, j, l} \, k_{T}^{ i}e_{2}^{j}e_{3}^{l} \, \delta_{p+3, i+2j+3l}
\\
&=\frac{1}{e_{3}^{3}k_{T}^{p}}\sum_{l=0}^{\floor{\frac{p+3}{3}}}\sum_{j=0}^{\floor{\frac{p+3-3l}{2}}}   C_{(p+3- 2j-3l), j, l} \, k_{T}^{(p+3- 2j-3l)}e_{2}^{j}e_{3}^{l} \,,
\end{align}
where $  C_{i,j,l} $ are numerical coefficients and the Kronecker delta ensures that the numerator has the correct degree. The symbol $ \floor{.}$ inicates the floor of its argument. This Ansatz is subject to two types of constraints. First, the residue of the leading $  k_{T} $ pole should correspond to an appropriate flat-space amplitude. If we restrict ourselves to the EFT of single-clock inflation, then we want to require that the corresponding amplitude is manifestly local, namely that it is a polynomial in $  e_{2} $ and $  e_{3} $. This constrains the leading term as $  k_{T}\to 0 $ to have at least one power of $  e_{3} $ in the numerator, and therefore $  C_{0i0}=0 $ for $  i>2 $ (the case $  C_{020} $ correspond to the not manifestly local amplitude generated by gravity that we discussed in the previous subsection). Second, the soft scalar theorem should be imposed to leading and next-to-leading order. This produces the constraints
\begin{align}
\sum_{j=0}^{\floor{\frac{p+3}{2}}} 2^{2-2j}\,C_{(p+3-2j),j,0}&=(1-n_{s})\,, \\ 
\sum_{j=0}^{\floor{\frac{p+3}{2}}} 4^{1-j}(3+2j) C_{(p+3-2j),j,0} +\sum_{j=0}^{\floor{\frac{p}{2}}} 4^{-j}\,C_{(p-2j),j,1} &=0 \,.
\end{align}
This gives the final bispectrum to $  p $-th order in derivatives in the EFT of inflation. All the remaining parameters correspond to linear combinations of the coupling constants in the theory. Their relative size is unconstrained by our current approach but it should be possible to include an additional rule that account for the existence of an approximate diagonal time-translation symmetry or whatever other symmetry is at play. \\

As an example, let us consider the most studied case of one derivative per field, as appropriate if the perturbations enjoy some (approximate) shift symmetry. For the bispectrum this gives a total of $  p=3 $ derivatives, and the Ansatz becomes
\begin{align}\label{any3}
B_{\zeta\zeta\zeta}=\left(  \frac{H^{2}}{4\e c_{s} \Mpl^{2}}\right)^{2}\frac{C_{0}e_{3}^{2}+\tilde C_{0}e_{2}^{3}+C_{1}k_{T}e_{2}e_{3}+C_{2}k_{T}^{2}e_{2}^{2}+C_{3}k_{T}^{3}e_{3}+C_{4}k_{T}^{4}e_{2}+C_{6}k_{T}^{6}}{e_{3}^{3}k_{T}^{3}}\,,
\end{align}
where we have simplified the notation by keeping only the first index in the free coefficients $  C_{i}=C_{i,j,l} $.
Manifest locality of the corresponding amplitude demands $  \tilde C_{0}= C_{0,3,0} =0 $. This avoids the appearance of the not manifestly local amplitude $ A_{3} \sim  e_{2}^{3}/e_{3} $. We are therefore left with $  7-1=6 $ free parameters. These parameters are further constrained by the leading (LO) and next-to-leading order (NLO) of the soft limit as in Bootstrap Rule 6:
\begin{align}\label{softl}
\frac{1}{4}(C_{2} + 4 C_{4} + 16 C_{6})&= 1-n_{s}\,, & (C_{1} + 7 C_{2} + 4 C_{3} + 20 C_{4} + 48 C_{6})&=0\,.
\end{align}
The next-to-leading order soft limit eliminates one parameter, while the leading order one trades another parameter for $  n_{s}-1 $. So the final result has $  6-1=5 $ free parameters, which can be taken to be $  \{1-n_{s},C_{0},C_{1},C_{2},C_{3}\} $. When computing the bispectrum explicitly in a generic $  P(X,\phi) $ model \cite{Chen:2006nt}, one finds also 5 free parameters at this order, namely $  \{c_{s},\lambda, s,\e,\eta\} $ where $  c_{s} $ is the speed of sound of scalar perturbations, $  s \equiv \dot c_{s}/(c_{s}H) $ parameterizes its dimensionless time variation, $  \lambda \equiv X^{2} P_{XX}+2X^{3}P_{XXX}/3$ and $  \e $ and $  \eta $ are the usual slow-roll parameters defined in \eqref{sr}. The two sets of parameter are related by\footnote{In principle one should be able to extract from 
\cite{Burrage:2011hd} the subleading slow-roll corrections that are left implicit in these formulae.}
\begin{align}
C_{0}&=18\left(  \frac{1-c_{s}^{2}}{c_{s}^{2}}-2\frac{\lambda c_{s}^{2}}{\e H^{2}}\right) \left[ 1+\O(\e,\eta,s) \right]+18s(3-2\gamma_{E})\,,\\
C_{1}&=-2 \left( \frac{1-c_{s}^{2}}{c_{s}^{2}} \right) \left[ 1+\O(\e,\eta,s) \right]+2s(2\gamma_{E}-3)\,,\\
C_{2}&=-2 \left( \frac{1-c_{s}^{2}}{c_{s}^{2}} \right) \left[ 1+\O(\e,\eta,s) \right]+2s(2\gamma_{E}-1)+4\e\,,\\
C_{3}&=\frac{11}{2} \left( \frac{1-c_{s}^{2}}{c_{s}^{2}} \right) \left[ 1+\O(\e,\eta,s) \right]-11(s\gamma_{E}+\e)+\frac{3}{2}\eta\,,\\
C_{4}&=-\frac{3}{2} \left( \frac{1-c_{s}^{2}}{c_{s}^{2}} \right)\left[ 1+\O(\e,\eta,s) \right]+s(1+3\gamma_{E})+2\e-\frac{3}{2}\eta\,,\\
C_{6}&=\frac{1}{2} \left( \frac{1-c_{s}^{2}}{c_{s}^{2}} \right) \left[ 1+\O(\e,\eta,s) \right]-s\gamma_{E}+\frac{1}{2}(\eta-\e)\,,
\end{align}
where $  \gamma_{E} $ is the Euler-Mascheroni constant. These expressions indeed satisfy the soft limit constraints  in \eqref{softl} with $  1-n_{s}=2\e+\eta+s $, which is the correct spectral tilt at this order. So the Bootstrap Rules were able to fully determine the EFT scalar bispectrum to this order. 

The main shortcoming of the current implementation of the Boostless Bootstrap is clear here: we can recover the correct bispectrum in terms of five free parameters, but we don't yet have a rule to tell us which of these parameters may be  large and which must be small. For example, it is known from explicit $  P(X,\phi) $ models or from the EFT of inflation that two of the free parameters above can be large, namely $  1/c_{s}^{2} $ and $  \lambda c_{s}^{2}/(\e H) $ (equivalent to $  M_{2} $ and $  M_{3} $ in the EFT of inflation). Linear combinations of these lead to the bispectrum templates known as equilateral and orthogonal non-Gaussianity. Conversely, other parameters must be much smaller than one, namely $  \e $, $  \eta $ and $  s $. In the Bootstrap derivation we can see a hint of this in the fact that both $  C_{0} $, which contains $  \lambda c_{s}^{2}/(\e H) $, and the specific linear combination of $  C $'s that corresponds to $  (1-c_{s}^{2})c_{s}^{-2} $ are not constrained by the squeezed limit \eqref{softl}. Conversely, $  \e $, $  \eta $ and $  s $, which must be small, also survive and contribute to the squeezed limit. It would be nice to find additional Bootstrap Rules that tell us which parameters can be large. This would presumably require enforcing the non-linear realization of Lorentz boosts, which provides the main constraining power in the structure of the EFT of inflation. We leave this for future investigation.

 
\subsection{One scalar and two graviton correlator $  \ex{\zeta\gamma\gamma} $}\label{ssec:ggz}

As a last example, let's see how to derive the mixed scalar-tensor-tensor bispectrum $  \ex{\gamma\gamma\zeta} $ from the Bootstrap Rules. We will only discuss the result for General Relativity (GR) minimally coupled to a canonical scalar field and postpone to a future publication a systematic derivation of all mixed correlators arising in the EFT of inflation (which, to the lowest orders in derivatives, were discussed in \cite{Creminelli:2014wna,Bordin:2017hal,Bordin:2020eui}).
This correlator displays characteristics that are similar to the scalar bispectrum, including a not manifestly local amplitude as the residue of the total energy pole \cite{Maldacena:2002vr}\footnote{The original result in \cite{Maldacena:2002vr} contained a typo in the coefficient of $  k_{1}^{3} $, which should be $  -1/2 $ as opposed to $  -1/4 $, as also noticed in \cite{McFadden:2011kk}. This can be checked by recomputing the contribution from the field redefinition in 3.18 of \cite{Maldacena:2002vr}. Alternatively, one may notice that with the incorrect factor of $  -1/4 $ the soft graviton limit does not vanish to LO, which it should.}
\begin{align}\label{zgg}
B_{\zeta\gamma\gamma}=\frac{H^{2}}{4\e \Mpl^{2}}\frac{H^{2}}{\Mpl^{2}} \frac{\e}{2} \frac{\e^{h_{2}}_{ij} \e^{h_{3}}_{ij}}{k_{1}^{3}k_{2}^{3}k_{3}^{3}}\left[  - \frac{1}{2} k_{1}^{3}+\frac{1}{2}k_{1}\left(  k_{2}^{2}+k_{3}^{2}\right)+4\frac{k_{2}^{2} k_{3}^{2}}{k_{T}} \right]\,.
\end{align}
To bootstrap this correlator, we begin by noticing that the only Lorentz invariant, manifestly local amplitude for one scalar (with momentum $ k_{1}  $) and two gravitons (with momenta $  k_{2} $ and $  k_{3} $) has four derivatives (corresponding to $  [23]^{4} $ in spinor helicity variables). Since we focus on a theory with at most two derivatives, the residue of the $  k_{T}^{-2} $ pole must vanish. As we did for the scalar bispectrum, the residue of the $  k_{T}^{-1} $ pole can be extracted from the flat space amplitude of a toy model with a massless scalar, a massless graviton and a constrained scalar, as discussed in Appendix \ref{app:toy}. This gives us the following amplitude (see \eqref{ggzamp})
\begin{align}\label{ampfixed}
A(1^{0},2^{h_{2}},3^{h_{3}})=i\frac{\dot \phi}{\Mpl^{2}}\e^{h_{2}}_{ij}(\v{k}_{2})\e^{h_{3}}_{ij}(\v{k}_{3})\frac{E_{2}E_{3}}{E_{1}}\,.
\end{align}
Note that this amplitude is not allowed in Minkowski spacetime because it does not obey consistent factorization \cite{Pajer:2020wnj}. However, it may be consistent in the presence of an IR modification, such as for example the curvature of spacetime in an FLRW universe. The most generic Ansatz that reproduces this amplitude is then
\begin{align}
B_{\zeta\gamma\gamma}&=\frac{H^4}{\Mpl^{4}}\frac{\e^{h_{2}}_{ij}(\v{k}_{2})\e^{h_{3}}_{ij}(\v{k}_{3})}{e_{3}^3 k_{T}^1} \left[   A_{4} k_{T}^4 + A_{3} k_{T}^3 (k_{2} + k_{3}) + A_{2} k_{T}^2 (k_{2} + k_{3})^2   \right.\\
   & \left . +B_{2} k_{T}^2 k_{2} k_{3} + A_{1} k_{T} (k_{2} + k_{3})^3 + 
   B_{1} k_{T} (k_{2} + k_{3}) k_{2} k_{3} +   \frac{1}{2}(k_{2} k_{3})^2\right]
\end{align}
where we used $  \Mpl H\sqrt{2\e}=\dot \phi $. The soft graviton theorem discussed in Bootstrap Rule 6 to linear order enforces
 \begin{align}\label{A4}
A_{1} + 2 A_{2} + 4 A_{3} + 8 A_{4}=0\,.
 \end{align}
The NLO term in the (symmetric) soft limit contains two terms: one proportional to the cosine of the angle $  \theta $ between the long and short modes and one that is independent of the angle. As discussed below \eqref{softg}, the spherical average NLO must vanish, which leads to (after using the solution of $  A_{4} $ in \eqref{A4})
\begin{align}
6 A_{1} + 10 A_{2} + 16 A_{3} + 24 A_{4} + 2 B_{1} + 4 B_{2}=0\,.
\end{align}
Using this to eliminate $  A_{1} $, the LO and NLO of the soft scalar limit enforce respectively
\begin{align}
-2 \frac{H^{4}}{3\Mpl^{4}}(-3 + 10 A_{2} + 16 A_{3} + 8 B_{1} + 22 B_{2}) &=\frac{H^{2}}{4\e \Mpl}\frac{H^{2}}{\Mpl}2\e \,, \\
 -1 + 6 A_{2} + 8 A_{3} + 2 B_{1} + 6 B_{2}&=0\,,
\end{align}
where we used \eqref{normeps}. Overall, solving these 4 constraints eliminates 4 variables and leaves 2, say $  A_{3} $ and $  B_{2} $ for concreteness. The part of the Ansatz containing free parameters is all non-singular in $  k_{T} $, since the only $  k_{T} $ pole was fixed to match the amplitude \eqref{ampfixed}. Terms that are finite as $  k_{T}\to 0 $ are related to boundary terms in time at the level of the Lagrangian and/or to field redefinitions. To discuss these terms, it is convenient to re-write them by reintroducing $  k_{1} $ 
\begin{align}
\frac{\Mpl^{4}} {H^4 e_{3}^{3}} B_{\zeta\gamma\gamma} &\supset  \e^{s_{2}}_{ij}(\v{k}_{2})\e^{s_{3}}_{ij}(\v{k}_{3}) \left\{   \frac{(k_{2} - k_{3})^2 (k_{2} + k_{3})}{224} (5 + 32 A_{3} + 88 B_{2})  \nonumber
 \right.\\
& \left. + \frac{k_{1}}{32}  \left[(1 - 2 B_{2}) \left( k_{2}^2 + k_{3}^2 \right) + 2 k_{2} k_{3} (1+ 8 B_{2} )\right]+\right.  \label{resca} \\
&\left.  + \frac{k_{1}^2 (k_{2} + k_{3})}{224} (-9 + 32 A_{3} - 24 B_{2})+  \frac{ k_{1}^3}{224} (-3 - 64 A_{3} - 8 B_{2}) \right\}\,.\nonumber
 \end{align}
Now we notice that $  k_{1} $ may not appear analytically in these terms. The reason is that the only field redefinitions that contribute to this correlator are those of $  \zeta $. Upon inspection, there are only two such independent field redefinitions that do not vanish at the boundary, namely $  \Delta \zeta \propto \gamma_{ij}\gamma_{ij} $ and $  \Delta \zeta \propto \partial^{-2}(\gamma_{ij}\partial^{2}\gamma_{ij}) $. Their contribution to the bispectrum may only contain integer powers of $  k_{1}^{2} $. Hence, when rescaled by $  e_{3}^{3} $ as in \eqref{resca}, $  \v{k}_{1} $ must appear non-analytically, e.g. as $  k_{1}=|\v{k}_{1}\cdot\v{k}_{1}|^{1/2} $ or $  k_{1}^{3} $, but not as $  k_{1}^{0} $ or $  k_{1}^{2} $. In particular, there are no field redefinitions of $  \gamma_{ij} $ that contribute to this correlator because it is not possible to write something at order $\O(  \zeta \gamma_{ij}) $ that maintains the transversality of $  \gamma_{ij} $. An equivalent way of thinking about this result is in terms of the hypothetical dual boundary CFT, whose partition function is the wavefunction of the universe in the bulk. In the boundary theory, one cannot have contact terms when the real-space positions of $  \zeta $ and $  \gamma_{ij} $ coincide. As a consequence, the momentum $  k_{1} $ of $  \zeta $ may not appear analytically in the Fourier-space wavefunction, which is precisely proportional to the right-hand side of \eqref{resca}. By imposing that the terms where $  k_{1} $ appears analytically in \eqref{resca} vanish, we find
\begin{align}
-9 + 32 A_{3} - 24 B_{2}&=0\,,& 5 + 32 A_{3} + 88 B_{2}&=0\,,
\end{align}
which we can solve to find the remaining parameters, $  A_{3}=3/16 $ and $  B_{2}=-1/8 $. Plugging this back in our Ansatz we finally find
\begin{align}
B_{\zeta\gamma\gamma}=\frac{H^{4}}{\Mpl^{4}}  \frac{\e^{s_{2}}_{ij} \e^{s_{3}}_{ij}}{k_{1}^{3}k_{2}^{3}k_{3}^{3}}\left[  - \frac{1}{16} k_{1}^{3}+\frac{1}{16}k_{1}\left(  k_{2}^{2}+k_{3}^{2}\right)+\frac{1}{2}\frac{k_{2}^{2} k_{3}^{2}}{k_{T}} \right]\,,
\end{align}
which indeed coincides with the result \eqref{zgg} of the explicit bulk calculation \cite{Maldacena:2002vr}.

 
\section{Conclusions}\label{sec:conc}

In this paper, we established a set of Bootstrap Rules that enforce symmetries, locality and the choice of vacuum on the bispectrum of massless scalars and gravitons around a quasi de Sitter spacetime. Then, we showed that many results in the literature as well as new ones follow from these rules. The general strategy is to start from a ``flat space'' amplitude\footnote{It should be noted that the amplitude that appears as the residue of the total-energy pole does not in general satisfy all the locality and unitarity constraints that apply to amplitudes in Minkowski. For example, boost breaking interactions are forbidden in Minkowski when coupling to a massless spin 2 field \cite{Pajer:2020wnj}. Furthermore, we stressed here that the amplitude associated to $ \ex{\zeta\gamma\gamma}  $ and to $  \ex{\zeta\zeta} $ have inverse powers of energy and therefore correspond to not manifestly local interactions.} and use the precise relation recently derived in \cite{COT} to fix the normalization of the correlator as well as its polarization-dependent part. Then, the assumption of working at tree-level in de Sitter spacetime with a Bunch-Davies initial state allows us to write down an Ansatz for the correlator with just a finite number of free numerical coefficients. The number of these coefficients grows with the total number of derivatives that one would like to include in the effective field theory derivative expansion, as expected. Finally, these free coefficients are constrained by soft theorems and symmetries. The remaining free coefficients correspond to linear combinations of coupling constants in the Lagrangian description. \\

The promising results derived here are an invitation to extend and generalize this approach. This can be done in several directions:
\begin{itemize}
\item The graviton bispectrum is constrained non-perturbatively by de Sitter isometries to be a linear combination of two possible terms \cite{Maldacena:2011nz}. In the presence of additional massive spin 2 fields, while still maintaining de Sitter isometries, additional interactions are allowed \cite{Goon:2018fyu}. However, in all models of inflation, de Sitter boosts are broken and this breaking is large for interactions that are produced by the coupling to the background foliation of time. This generates an infinite set of new correlators involving the graviton, including additional graviton non-Gaussianities. These have been studied to leading order in derivatives \cite{Creminelli:2014wna,Bordin:2017hal,Bordin:2020eui}. It would be interesting to find a general expression to any order in derivatives for spinning correlators, analogous to the general expression we found here for scalars, \eqref{any}.
\item In this work we did not invoke unitarity explicitly. The only rule where some information about unitarity might be implicitly hiding is Bootstrap Rule 6 about soft theorems. Conversely, four- and higher-point correlation functions involving exchange diagrams are very strongly constrained by unitarity. To extend the Boostless Bootstrap to exchange correlators we need therefore a new ingredient. The Cosmological Optical Theorem (COT) recently derived in \cite{COT} and further discussed in \cite{Cespedes:2020xqq} is probably that ingredient (see also \cite{Meltzer:2020qbr} for parallel developments on the AdS side). The COT is a constraint on the analytical continuation of boundary correlators, and the corresponding wavefunction coefficients, that is dictated by unitary evolution in the bulk. The COT determines four- and higher-point correlators/wavefunction coefficients in terms of lower ones up to ``contact'' terms, namely terms with a vanishing right-hand side. It would be interesting to use the bispectra derived here to bootstrap trispectra using the COT.
\item It would be advantageous to exploit the spinor helicity formalism to better understand the analytic structure of correlators and the wavefunction in the presence of spinning fields. This approach should be particularly interesting in the presence of massless particles, just as it is the case in flat spacetime. 
\item In the study of amplitudes in Minkowski, the role of locality at tree level is well understood and encoded in the consistent factorization. However, when boosts are non-linearly realized, the implications of locality become important already at the level of contact interactions. Indeed, we emphasized here that both $  \ex{\zeta\zeta\zeta} $ and $  \ex{\zeta\gamma\gamma} $ have not manifestly local amplitudes appearing on their total-energy poles. As expected since we are dealing with gravity, these amplitudes and the corresponding correlators are local in the sense defined in Rule 5. Further interesting models of non-manifest locality arise in several of the possible ways in which boosts can be broken \cite{Zoology}. Furthermore, saturating locality constraints should be related to the existence of interacting massless particles, which can mediate long range forces. Given the prominent role that locality plays in fundamental physics and in on-shell approaches, it is important to develop a more thorough and coherent understanding of the observations above.
\end{itemize}
While a few aspects of the current approach are still not completely satisfactory, we believe that our findings are very encouraging and one should be able to develop a more systematic and robust understanding of cosmological correlators from the boundary without any reference to de Sitter boosts. More ambitiously, we may hope and speculate that the rules we found here and their generalizations and extensions might one day play the same role that the ad hoc rules in Bohr's atomic model played for quantum mechanics. Eventually, all these rules should be understood as elements of a more fundamental notion of de Sitter holography and a new perspective on the notion of time in quantum theories of gravity.

 
\section*{Acknowledgments}\label{sec:} 

There have been many interesting discussions over the past two years that have influenced the ideas presented here. I am particularly thankful to Nima Arkani-Hamed, Daniel Baumann, Paolo Benincasa, Giovanni Cabass, Harry Goodhew, Tanguy Grall, Daniel Green, Maria Alegria Gutierrez, Aaron Hillman, Sadra Jazayeri, Austin Joyce, Scott Melville, Guilherme Pimentel, David Stefanyszyn and Jakub Supe\l. I would also like to thank the organizers and the participants of the \href{https://www.simonsfoundation.org/event/amplitudes-meet-cosmology-2019/29}{``Amplitudes meet Cosmology'' workshop}, where this project started. The author has been supported in part by the research program VIDI with Project No. 680-47-535, which is (partly) financed by the Netherlands Organization for Scientific Research (NWO).


\appendix

\section{Irreducible representations of ISO(3)}\label{irreps}

In this section, we classify cosmological perturbations around FLRW according to the irreducible representations of the Euclidean group. The discussion follows closely the one in  Chapter 2 of \cite{Weinberg:1995}. To find the irreps of ISO(3) we need to find a set of matrices $ U(R,\alpha)   $ for each ISO(3) element $  \{R^{i}_{j},\alpha_{l}\} $ that act on some Hilbert (vector) space of perturbations. In the following I will borrow the language from Quantum mechanics and refer to perturbations as ``states'' or ``state-vectors''. To begin, we note that ``the components of the three-momentum all commute with each other and so it is natural to express physical state-vectors in terms of eigenvectors of the three-momentum.'' \cite{Weinberg:1995}. This is the usual Fourier transform: we consider state-vectors that are eigen-functions of translations
\be
\hat P^{i} \psi_{k\sigma}=k^{i}\psi_{k\sigma}\,,
\ee 
where $  \sigma $ is some other (discrete) quantum number that we have to figure out. Translations are represented by the unitary transformation 
\be\label{transl}
U(1,\alpha)\psi_{k\sigma}=e^{-ik^{i}\alpha_{i}}\psi_{k\sigma}\,.
\ee
Now, we want to find the action of rotations $  U(R,0)\equiv U(R) $. Using the group properties, we note that
\be
U(R) \psi_{k\sigma}=C_{\sigma\sigma'}(R,k)\psi_{Rk\sigma'}\,,
\ee
that is, a rotation changes the three-momentum of the state. We want now to find irreducible $  C_{\sigma \sigma'} $ (i.e. that cannot be decomposed into smaller blocks by changing the basis for $  \psi_{k\sigma} $). For this we will use the method of induced representations. The subgroup of ISO(3) we will be interested in is SO(3). The only invariant under SO(3) is the norm of a vector (and any function thereof), $  k^{i}k^{j}\delta_{ij}=k^{2} $. Let us play some algebraic tricks now. For a reference vector $   q^{i} $, define the rotation $ S(k) $ that transforms it into any other vector $  k^{i} $ as
\be
S(k) q = k \then S^{-1}(k)k=q\,.
\ee
We can then re-write any state with momentum $  k $ as a transformation of a state with reference momentum $  q $,
\be
\psi_{k\sigma}=U(S(k))\psi_{q\sigma}\,.
\ee
Then, the action of a general rotation $  R $ can be massaged as follows:
\be
U(R) \psi_{k\sigma}&=& U(R)U(S(k))\psi_{q\sigma}\\
&=&U(S(Rk))U(S^{-1}(Rk)RS(k)) \psi_{q\sigma}\\
&=&U(S(Rk)) D_{\sigma\sigma'}\psi_{q\sigma'}\\
&=&D_{\sigma\sigma'} U(S(Rk)) \psi_{q\sigma'}\\
&=&D_{\sigma\sigma'} \psi_{Rk\sigma'}\,,\\
\ee
where in the third line we recognised that $ S^{-1}(Rk)RS(k) q=q  $ and so 
\be
U(S^{-1}(Rk)RS(k)) \psi_{q\sigma}\equiv D_{\sigma\sigma'}\psi_{q\sigma}\,,
\ee
i.e. it must be some linear combination $  D_{\sigma\sigma'} $ of states with momentum $  q $. From this definition of $  D_{\sigma\sigma'} $, we see that is it provides a representation of the little group, namely the subgroup of SO(3) that leaves the representative vector $  q $ invariant. For every little group rotation $  r $, we have
\be
U(r)\psi_{q\sigma}=D_{\sigma\sigma'}(r)\psi_{q\sigma'}\,.
\ee
Summarising, choosing a representative vector $  q $ and given a representation $  D_{\sigma\sigma'}  $ of the little group for $  q $, we get a representation of the full group ISO(3) defined by
\be
\boxed{\begin{array}{rcl} U(1,\alpha)\psi_{k\sigma}&=& e^{-ik^{i}\alpha_{i}}\psi_{k\sigma}\,,\\
U(R,0)\psi_{k\sigma}&=& D_{\sigma\sigma'}(r(R,k)) \psi_{Rk\sigma'}\,, \end{array}}
\ee
where the little group element $  r(R,k) $ is given by 
\be
r(R,k)\equiv S^{-1}(Rk)RS(k)\,.
\ee

 
\subsection*{Little groups}\label{ssec:}

While for the Poincar\'e group there are 6 little groups, of which only three have physical significance (the vacuum, massive particles and massless particles), for cosmology there are only two little groups: SO(3) itself for $  q^{i}q_{i}=0 $, and SO(2) for $  q^{i}q_{i}\neq 0 $. 

The irreps of SO(3) are well known from the study of angular momentum in quantum mechanics. They are classified by the Casimir operator $  J^{2} $, with eigen-values $  l(l+1) $ for $  l=0,1/2,1,\dots $ and are of dimension $  2l+1 $ with states $  \ket{l,m} $ and $ |m|\leq l $. Focussing on the bosonic irreps with integer $  l $, we know they correspond to spin zero, one, two, etc. The field operators that generate those states are:
\be
\text{Spin zero:}&&\quad \phi,\,h_{ii},\,\dots\\
\text{Spin one}&& \quad h_{0i},\,u_{i},\,\dots\\
\text{Spin two:}&&\quad h_{\ex{ij}}\equiv h_{ij}-\frac{1}{3}h_{kk}\delta_{ij},\,\dots\,.
\ee
Notice that the splitting between the trace of the two-tensor $  h_{ij} $, which has spin zero, and its traceless part $  h_{\ex{ij}} $, which has spin two, is purely algebraic and does not involve any (inverse) Laplacians. These $  q=0 $ irreps are relevant to classify and discuss the background and adiabatic modes. For physical perturbations, we have to consider the other representative vector.

For $  q^{i}q_{i}\neq 0 $, we can choose as representative vector $  q^{i}=\{q,0,0\} $ so that the little group is recognised as two-dimensional rotations, namely SO(2), which is an abelian group. All complex representations of an Abelian group are one-dimensional by Schur's lemma (all real representations are two dimensional). There are infinitely many such representations, enumerated by an integer $  m\in \mathbb {N} $. Physically, we can interpret $  m $ as the ``helicity'' of the state, i.e. how it transforms under a rotation around the direction of its momentum. If the underlying theory is parity invariant, which is sometimes assumed in cosmological applications, for every state with helicity $  m $ there as to exist a state of helicity $  -m $. So we have classify states as helicity 0, 1, 2 etc.


\section{Apparent non-locality from a non-dynamical field} \label{app:toy}

In this appendix, we present a toy model of how the presence of the non-dynamical lapse and shift in the calculation of the scalar bispectrum from gravitation interaction leads to apparent non-locality \cite{Maldacena:2002vr}. While we partially use the amplitude language to tell this story, we explicitly resort to a toy Lagrangian to make the discussion most explicit and familiar. We start with a scalar amplitude and then include gravitons as well.\\

\paragraph{Scalar theory} Consider the simple ``local'' but non-Lorentz invariant Lagrangian\footnote{The normalization of the $  \chi$-$\phi $ interaction is obtained by replacing $  \gamma_{\mu\nu} $ with $  \chi_{NC} \delta_{\mu0}\delta_{\nu0} $ in the minimal coupling to gravity, where $  \chi_{NC} $ is the non-canonically normalized version of $ \chi  $,
\begin{align}
\L&=-\frac{1}{2}\left( \partial_{\mu}\phi \right)^{2}+\frac{M^{2}}{4}\left( \partial_{i}\chi_{NC} \right)^{2}+\frac{1}{2}\chi_{NC}\dot\phi^{2}-\tilde\Lambda^{3}\chi_{NC}\,.
\end{align}}
\begin{align}\label{orig}
\L&=-\frac{1}{2}\left( \partial_{\mu}\phi \right)^{2}-\frac{\Mpl^{2}}{4}\left( \partial_{i}\tilde \chi \right)^{2}+\frac{1}{2}\chi \dot\phi^{2}-\tilde\Lambda^{4}\chi\,,\\
&=-\frac{1}{2}\left( \partial_{\mu}\phi \right)^{2}-\frac{1}{2}\left( \partial_{i}\chi \right)^{2}+\frac{1}{\sqrt{2}\Mpl}\chi \dot\phi^{2}-\Lambda^{3}\chi\,,
\end{align}
where $  \phi $ is a canonical scalar field derivatively coupled to a non-dynamical, canonically-normalized scalar $  \chi $ (related to the non-canonical $  \tilde \chi $ by $   \chi=\Mpl \tilde\chi/\sqrt{2} $), which plays the role of the lapse. The tadpole for $\chi  $ has been introduced just to get a suitable classical background and does not play a role at the perturbative quantum level. We would like to expand around some non-trivial homogeneous background $  \phi=\bar \phi +\varphi $. The homogeneous equations of motion in Minkowski give us
\begin{align}
\ddot \phi \left(  1+\frac{2}{M}\chi \right)+\frac{\sqrt{2}}{\Mpl}\dot \phi \dot \chi &=0\,, \\
\frac{1}{\sqrt{2} \Mpl}\dot{\bar \phi}^{2}-\Lambda^{3}&=0\,,
\end{align} 
which is solved by $  \chi = 0 $ and $  \phi =t \dot{\bar \phi} $, with $  \dot{\bar \phi}=2^{1/4}\sqrt{\Mpl\Lambda^{3}} $ and the missing integration constants were chosen to vanish for convenience.
Now we would like to compute the tree-level three-$  \varphi $ amplitude. We can do this in three ways:
%

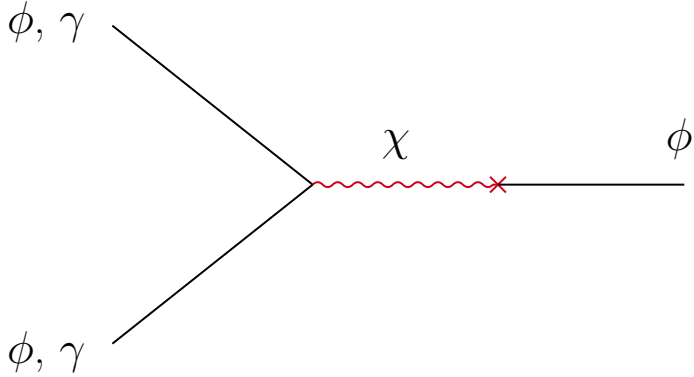
\begin{figure}
\centering
\tikzset{every picture/.style={line width=0.75pt}} 

\begin{tikzpicture}[x=0.75pt,y=0.75pt,yscale=-1,xscale=1]

\draw    (61,40) -- (161,120) ;
\draw [color={rgb, 255:red, 208; green, 2; blue, 27 }  ,draw opacity=1 ]
   (161,120) .. controls (162.67,118.33) and (164.33,118.33) .. (166,120)
.. controls (167.67,121.67) and (169.33,121.67) .. (171,120) .. controls
(172.67,118.33) and (174.33,118.33) .. (176,120) .. controls
(177.67,121.67) and (179.33,121.67) .. (181,120) .. controls
(182.67,118.33) and (184.33,118.33) .. (186,120) .. controls
(187.67,121.67) and (189.33,121.67) .. (191,120) .. controls
(192.67,118.33) and (194.33,118.33) .. (196,120) .. controls
(197.67,121.67) and (199.33,121.67) .. (201,120) .. controls
(202.67,118.33) and (204.33,118.33) .. (206,120) .. controls
(207.67,121.67) and (209.33,121.67) .. (211,120) .. controls
(212.67,118.33) and (214.33,118.33) .. (216,120) .. controls
(217.67,121.67) and (219.33,121.67) .. (221,120) .. controls
(222.67,118.33) and (224.33,118.33) .. (226,120) .. controls
(227.67,121.67) and (229.33,121.67) .. (231,120) .. controls
(232.67,118.33) and (234.33,118.33) .. (236,120) .. controls
(237.67,121.67) and (239.33,121.67) .. (241,120) .. controls
(242.67,118.33) and (244.33,118.33) .. (246,120) .. controls
(247.67,121.67) and (249.33,121.67) .. (251,120) -- (253.5,120) --
(253.5,120) ;
\draw [shift={(253.5,120)}, rotate = 45] [color={rgb, 255:red, 208;
green, 2; blue, 27 }  ,draw opacity=1 ][line width=0.75]    (-5.59,0) --
(5.59,0)(0,5.59) -- (0,-5.59)   ;
\draw    (61,200) -- (161,120) ;
\draw    (253.5,120) -- (346.5,120) ;

\draw (38,192.4) node [anchor=north west][inner sep=0.75pt] 
[font=\LARGE]  {\hspace{-1cm} $\phi $, $  \gamma $};
\draw (38,26.4) node [anchor=north west][inner sep=0.75pt]  [font=\LARGE]
  {\hspace{-1cm} $\phi $, $  \gamma $};
\draw (194,85) node [anchor=north west][inner sep=0.75pt]  [font=\LARGE]
[align=left] {$\displaystyle \chi $ \hspace{3cm} $\displaystyle
\phi $};
\end{tikzpicture}
\caption{\label{Fey} The Feynman diagram for the $  \varphi\varphi\varphi $ and $  \gamma\gamma\varphi $ amplitudes mediated by the massless, non-dynamical scalar field $  \chi $.}
\end{figure}
\begin{enumerate}
\item Integrate out $  \chi $ at tree-level in perturbation theory. Its equations of motion are
\begin{align}
\partial_{i}^{2}\chi=\frac{\sqrt{2}}{\Mpl}\dot{\bar \phi}\dot \varphi +\dots\,,
\end{align}
where we can neglect the quadratic term in $  \varphi $. Plugging this back into the Lagrangian we find
\begin{align}
\L = \frac{1}{2}\partial_{\mu}\varphi^{2}+\frac{\dot{\bar \phi}}{\Mpl^{2}}\dot \varphi^{2} \partial_{i}^{-2}\dot \varphi\,.
\end{align} 
Not surprisingly, integrating out a massless (non-dynamical) field has generated some form of non-locality. Furthermore, the non-local cubic interaction above is precisely the interaction leading to the amplitude
\begin{align}
A_{\varphi\varphi\varphi}&= 2 i \frac{\dot{\bar\phi}}{\Mpl^{2}} E_{1}E_{2}E_{3}\left(  \frac{1}{E_{1}^{2}}+ \frac{1}{E_{2}^{2}}+ \frac{1}{E_{3}^{2}}\right)\\
&= 2 i \frac{\dot{\bar\phi}}{\Mpl^{2}}\frac{e_{2}^{2}}{e_{3}}\,,
\end{align}
where in the last line we used conservation of energy $  E_{T}=0 $. This amplitude has the correct mass dimension to correspond to $  p=1 $, namely $  2\times 2 - 3=1 $ overall derivative. 
\item We can keep $  \chi $ and derive the relevant Feynman rules directly for \eqref{orig}, which working perturbatively in $  \Mpl^{-1} $ will include a quadratic mixing $ \sqrt{2} \dot{\bar \phi} \chi \dot \varphi /\Mpl $. The Feynman diagram in Figure \ref{Fey} then gives again the same amplitude
\begin{align}
A_{\varphi\varphi\varphi}&=-i\times 2i\frac{E_{1}E_{2}}{\sqrt{2}\Mpl} \times \frac{i}{|\v{k}_{3}|^{2}} \times \frac{\sqrt{2}\dot{\bar \phi}}{\Mpl} E_{3} +\left( \text{2 perm's}  \right)\\
&=2i \frac{\dot{\bar \phi}}{\Mpl^{2}}E_{1}E_{2}E_{3}\left(  \frac{1}{E_{1}^{2}}+ \frac{1}{E_{2}^{2}}+ \frac{1}{E_{3}^{2}}\right)\\
&=2i \frac{\dot{\bar \phi}}{\Mpl^{2}} \frac{e_{2}^{2}}{e_{3}}\,.
\end{align}
\item We can derive $  A_{\varphi\varphi\varphi} $ from a two-to-two scattering amplitude by ``putting a leg on the background''. The four-$  \varphi $ scattering amplitude mediated by $  \chi $ is
\begin{align}\label{A4app}
A_{4}&= (-i) \times2 (-i) \frac{E_{1}E_{2}}{\sqrt{2}\Mpl}\times \frac{i}{|\v{k}_{1}+\v{k}_{2}|^{2}}\times 2 (-i) \frac{E_{3}E_{4}}{\sqrt{2}\Mpl}+\text{2 perm's}\,.
\end{align}
The cubic amplitude then is obtained rescaling by a factor $   \dot{\bar \phi}/(iE_{4}) $, which substitutes $  \dot \varphi $ with $  \dot{\bar \phi} $ and sending one leg to zero, say $ E_{4} , \v{k}_{4}\to 0 $:
\begin{align}
A_{\varphi\varphi\varphi}&= \lim_{E_{4},\v{k}_{4}\to 0}\, \frac{\dot{\bar \phi} }{iE_{4}}A_{4}= 2i\frac{\dot{\bar \phi}}{\Mpl^{2}}E_{1}E_{2}E_{3}\left(  \frac{1}{E_{3}^{2}}+ \frac{1}{E_{2}^{2}}+ \frac{1}{E_{1}^{2}}\right)=2i \frac{\dot{\bar \phi}}{\Mpl^{2}} \frac{e_{2}^{2}}{e_{3}}\,.
\end{align}
\end{enumerate}

\paragraph{Scalar-tensor theory} We will now proceed similarly to the scalar case, but add gravitons into the discussion. Our final goal is to derive the apparently non-local amplitude for two gravitons and one scalar, which appears on the residue of the $  k_{T}^{-1} $ of the corresponding cosmological correlator $  \ex{\zeta\gamma\gamma} $. We start from the toy model
\begin{align}\label{orig}
\L&=-\frac{1}{2}( \partial_{\mu}\phi)^{2}+\frac{\Mpl^{2}}{8} \left[ - 2(\partial_{i}\tilde\chi)^{2}-(\partial_{\mu}\tilde{\gamma}_{ij})^{2}+\tilde{\chi}   \dot{\tilde{\gamma}}_{ij}^{2}  \right] +\frac{1}{2}\tilde{\chi} \dot \phi^{2}-\tilde \Lambda^{4}\tilde\chi \\
&=-\frac{1}{2}( \partial_{\mu}\phi)^{2} - \frac{1}{2}(\partial_{i}\chi)^{2}+\frac{1}{\sqrt{2} \Mpl}\chi \dot\phi^{2}-\frac{1}{4}(\partial_{\mu}\gamma_{ij})^{2}+\frac{1}{2\sqrt{2}\Mpl} \chi \dot \gamma_{ij}^{2} -\Lambda^{3}\chi\,.
\end{align}
There is a similar diagram as in Figure \ref{Fey} with the two left $  \phi $ legs substituted by two $  \gamma $ legs. The three-particle amplitude for this diagram, namely for one scalar and two gravitons that is mediated by the non-dynamical field $  \chi $ is found to be
\begin{align}
A_{\varphi\gamma\gamma}&=(-i)\times (-i)2 \frac{iE_{2}iE_{3} \e_{ij}^{s_{2}}\e_{ij}^{s_{3}}}{2\sqrt{2}\Mpl}\times\frac{i}{|k_{1}|^{2}}\times (-i)\frac{2\dot\phi}{\sqrt{2}\Mpl} iE_{1}\\
&=i\frac{\dot\phi}{\Mpl^{2}}\e_{ij}^{s_{2}}\e_{ij}^{s_{3}} \frac{E_{2}E_{3}}{E_{1}}\,.\label{ggzamp}
\end{align}

\bibliographystyle{utphys}
\bibliography{refs}

\end{document}